# Theoretical Study on the Potential Existing Forms and Microwave Rotational Spectrum of Short-Chain Fatty Acids in Interstellar Space

Fangjing Mu[1], Hao Wang[2], Zhanhang He[3], Qian Gou[2,*], Yuchao Zhang[4,*], and Donghui Wei[3,*]

[1] Academy for Advanced Interdisciplinary Studies (AAIS), Peking University, Beijing 100871, China.
[2] Green Catalysis Center, and College of Chemistry, Zhengzhou University, Zhengzhou, Henan 450001, China.
[3] Department of Chemistry, School of Chemistry and Chemical Engineering, Chongqing University, Chongqing 401331, China
[4] The Cancer Hospital of the University of Chinese Academy of Sciences, Institute of Basic Medicine and Cancer (IBMC), Chinese Academy of Sciences, Hangzhou, Zhejiang 310022, China
*Email: qian.gou@cqu.edu.cn (Qian Gou), yuchaozhang@ucas.ac.cn (Yuchao Zhang), donghuiwei@zzu.edu.cn (Donghui Wei)

**Abstract**
Several short-chain fatty acids and their corresponding potential existing hydrated forms are important molecules in interstellar space. Their structures were optimized with twelve different computational methods. The dipole moments and the spectral constants, including rotational constants and centrifugal distortion constants were obtained. According to the benchmark study, revDSD-PBEP86-D3(BJ) is the most suitable method that was selected for rotational calculation. Symmetry-adapted perturbation theory was used to study the strength and composition of the interaction between acids and water in clusters. The possibility of its existing under the low-temperature and low-pressure conditions was confirmed by calculating of binding free energy. Furthermore, *ab initio* molecular dynamics simulations were used to investigate whether the internal rotations of acids could be observed. The 3-fold splitting from the predicted high-resolution microwave rotational spectra of the acetic acid monohydrate at different temperatures perfectly proved the accuracy of the simulations.

## 1. Introduction

Microwave Rotational Spectra are used to not only diagnose the physical state of molecules in circumstellar nebulae and envelopes., but also confirm the molecules in interstellar space, then the abundance of molecules can be reflected by strength of spectra, thus indicating the temperature of the environment. Derived from this, organic molecules such as formic acid and acetic acid have been widely observed in space by assistance of its microwave rotational spectrum[1]. Short-chain fatty acids (SCFAs), also known as volatile fatty acids, usually refer to saturated fatty acids with 1 to 6 carbon atoms. Apart from being observed by telescope, they have also been found in meteorites. Propionic acid is the most abundant monocarboxylic-acid-type soluble organic matter in Murchison meteorites[2–4]. Generally, fatty acids play a pivotal role in the origin of life because SCFAs and their hydrates are the basis for the existence of carbon-based life. SCFAs are precursor molecules of life and form the basis for the existence of carbon-based life, and have significant biological significance. It's putative that fatty acids form biological membranes by combining with glycerol and phosphoric acid to form phospholipids. Noteworthy, fatty acid membranes bind to amino acids, which can stabilize biomembranes and they can contribute to the eventual formation of proteins[5]. In addition, SCFAs are the main metabolic products by microbiota[6], in which case, they play an important role in physiological and biochemical processes[7]. Consequently, it is possible to study the composition of the upper atmosphere of a planet by studying the microwave spectral satellite lines of organic acids and their water clusters, and then to search for life matter and explore whether extraterrestrial life exists in the vast universe.

Recently, tremendous advances have been achieved in the study of microwave rotational spectroscopy of small organic molecules. The widely use of Fourier transform microwave spectrometer provide great convenient to scientists to obtained multifarious rotational spectra of organic molecules on the earth, and extracted more accurate structure information. In order to investigate the configuration of SCFAs hydrate clusters, the Howard group studied the microwave spectra of propionic acid (PPA) and its hydrates by configuring solutions with different ratios in 2008, obtaining information such as molecular structure, spectroscopic constants, and binding energy[8]. Similarly, in 2009, the group studied the microwave spectra of acetic acid (AA) and its hydrate clusters[9], and it can be inferred that SCFAs hydrate clusters may exist in the form of monohydrates and dihydrates in interstellar space. Weak interactions exist in hydrate clusters, and in experiments, conformational analysis and hydrogen bond parameters can be studied by analyzing the microwave rotational spectra[10]. In 2017, Caminati and his coauthors analyzed the rotation of acetic acid and fluorinated acetic acid dimers, revealing the role of hydrogen bonding between organic acids[11]. Additionally, Internal rotation in interstellar molecules is an important research topic in astrochemistry. Experimental scientists can observe the microwave rotational spectra of small molecules using instruments such as FTMW, and then analyze internal rotation and other large-amplitude motions through spectral line splitting. In 2011, Caminati's research group studied the rotational spectra of 2-butyric acid, mainly exploring internal rotation issues[12]. In 2017, the group analyzed the rotational motion of acetic acid and fluorinated acetic acid dimers[13]. To the best of our knowledge, systematic theoretical computational studies on the microwave rotational spectroscopic properties of common volatile SCFAs and their hydrated clusters lag behind.

In this work, we aim to simulate straight SCFAs (Ac) and their hydrate clusters, namely acetic acid (AA), propionic acid (PPA), butyric acid (BA), pentanoic acid (PTA), hexanoic acid (HXA) and their monohydrate clusters (Ac-w) and dihydrate clusters (Ac-$w_2$). We only consider mono- and dihydrates because the collision and combination probability of SCFAs with three or more molecules of water has already been relatively low. Systematic calculations were performed on them using *ab initio* and DFT methods. Information like geometric structures, rotational constants ($A$, $B$, $C$), centrifugal distortion constants ($\Delta_J$, $\Delta_K$, $\Delta_{JK}$, $\delta_J$, $\delta_K$), molecular electric dipole moments, and components in orthogonal directions ($|\mu_{TOT}|$, $|\mu_a|$, $|\mu_b|$, $|\mu_c|$), isomerization transition state and corresponding Gibbs free energy barrier ($\Delta G^{\#}$) of internal rotations can be obtained. It should be mentioned that the most suitable calculation method for such system was confirmed after being compared with the existing experimental results. Additionally, the composition of dissociation energy ($\Delta G$) was studied and the internal rotation and cluster stability were simultaneously studied using *ab initio* molecular dynamics (AIMD) innovatively at different temperatures. These results will provide new insights for the development of molecular spectroscopy and its application in the chemistry field.

## 2. Computational details

In the very first place, we constructed structural models of AA, PPA, BA, PTA, HXA and their mono- and dihydrates, and performed the completely geometric optimization using the Gaussian16 package[14]. Applied with Grimme's D3 correction[15] with Becke-Johnson (BJ)[16], we used Generalized Gradient Approximation (GGA) BLYP-D3(BJ)[17–19], meta-Generalized Gradient Approximation (meta-GGA) MN15L-D3[20,21], Hybrid GGAs B3LYP-D3(BJ)[22–24], PBE0-D3(BJ)[25–27] and $\omega$B97X-D[28], Hybrid meta-GGAs M06-2X(D3)[29,30] and MN15-D3(BJ)[21,31], Double Hybrid Density Functionals (DHDFs) B2PLYP-D3(BJ)[32,33], spin-scaling DHDFs DSD-PBEP86-D3(BJ)[34] and revDSD-PBEP86-D3(BJ)[35], and wave function (WF) based methods HF and MP2, which are combined with basis set may-cc-pVTZ[36–38] for all atoms, respectively. The optimized structures were confirmed to be minimum points by frequency analyses, and the Cartesian coordinates were given in **Supporting Information**. Since they are all asymmetric-top molecules, the dipole moment obtained by structural optimization can determine the type of rotational spectra. In addition, the centrifugal distortion constants were acquired by calculating the resonant frequency of the vibration-rotation coupling. Compared with the calculated rotational constants of AA, PPA and their hydrates with the experimental values[8,9], the optimal calculation method for this kind of system was identified. Furthermore, the Root Mean Square Difference (RMSD) was calculated based on the geometric optimization results, and different methods were evaluated to reflect the reliability of optimizing the structures.

Applying high order symmetry-adapted perturbation theory (SAPT)[39], we used the PSI4 1.5[40] program to perform the energy decomposition calculation for the weak interaction in the SCFA hydrates model at the gold standard[41,42] of SAPT2+(3)δMP2/aug-cc-pVTZ[38,43–45] with unfrozen nucleus, and the total interaction energy and its four energy components were obtained. Moreover, the dissociation Gibbs free energies at different temperatures were corrected using Shermo 2.3[46] software from the results above. Including dispersion interactions, we employed the revDSD-PBEP86-D3(BJ) functional combined with may-cc-pVTZ basis set to scan the rotation of terminal methyl dihedral ($\Phi$), and then located the rotation transition state and calculated the $V_3$ energy barrier for terminal methyl rotation. Ultimately, we performed AIMD studies derived from adiabatic approximation for acetic acid monohydrate (AA-w), butyric acid (BA-w), and (HXA-w) using ORCA 5.0.2 program[47,48] with PBE0-D3(BJ)/def2-TZVP[49], and further got the stability and vibration characteristics of the clusters within 1 ps time scale at 5 K, 50 K, and 100 K, separately. Meanwhile, Rijcosx algorithm [50] was used to accelerate the calculation by reducing the time-consuming calculation of hybrid functionals dramatically. Predicted high resolution microwave rotational spectra were used to study gaseous molecules in isolated systems under vacuum, and they were obtained from Herb Pickett's SPCAT program[51]. All of the structures were visualized by Visual Molecular Dynamics (VMD) software[52] and the microwave spectra were visualized by pgopher[53].

## 3. Results and discussion

3.1 Optimized structures

Inspired by the conformations of AA hydrates[54], we have constructed the a series of initial structures of Ac, Ac-w and Ac-$w_2$. We have optimized structures of the various acids Ac and the corresponding water clusters Ac-w and Ac-$w_2$ at the twelve computational levels, and those structures optimized at the revDSD-PBEP86-D3(BJ)/may-cc-pVTZ level have been summarized in **Table 1**. From the average of the RMSD values provided in **Table S1** of **SI**, the differences between the geometrical structures are tiny, indicating the optimized structures are reliable.

**Table 1** The optimized structures of the Ac and the corresponding Ac-w and Ac-$w_2$ at the revDSD-PBEP86-D3(BJ)/may-cc-pVTZ level

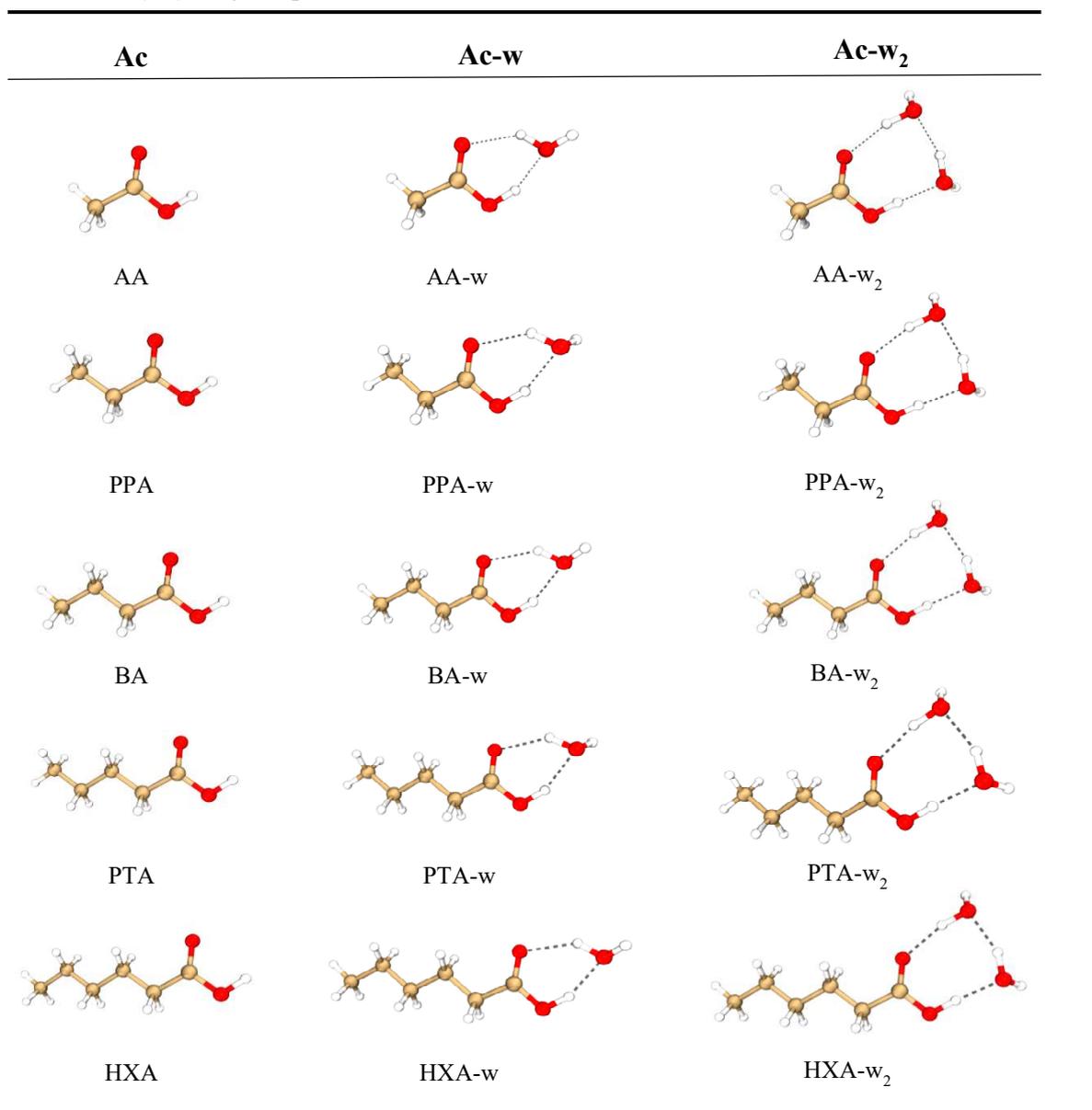

On the basis of the optimized structure, the same level calculations were performed to obtain the microwave rotational spectral constants, including the rotational constants and centrifugal distortion constants, for all the Ac, Ac-w, Ac-w2 by using different methods, which can be found in **Table 2** and **Tables S2.1-2.14** of **Supporting Information (SI)**. Our calculation results of AA and PPA summarized in **SI** are similar to those reported by Howard and Stiefvater et al. before[8,9,55], indicating that calculation results are reasonable and reliable.

**Table 2** Rotational constants, centrifugal distortion constants, and dipole moment of PTA-w calculated by using the twelve different methods

| PTA-w | B2PLYP-D3(BJ) | B3LYP-D3(BJ) | BLYP-D3(BJ) | DSD-PBEP86-D3(BJ) | HF | M06-2X(D3) | MN15-D3(BJ) | MN15L(D3) | MP2 | PBE0-D3(BJ) | revDSD-PBEP86-D3(BJ) | $\omega$B97X-D |
|---|---|---|---|---|---|---|---|---|---|---|---|---|
| $A$ (MHz) | 6265.7469 | 6282.6554 | 6178.8430 | 6251.2776 | 6354.6294 | 6258.7835 | 6232.8379 | 6075.0504 | 6244.8708 | 6338.1422 | 6243.7290 | 6317.8185 |
| $B$ (MHz) | 636.9897 | 635.2661 | 625.9084 | 638.9130 | 617.2387 | 639.6434 | 641.3249 | 620.6068 | 640.9577 | 644.3143 | 636.3199 | 635.6219 |
| $C$ (MHz) | 587.1183 | 585.8340 | 577.1507 | 588.6847 | 570.7288 | 589.2361 | 590.4707 | 571.6824 | 590.3146 | 594.0652 | 586.4236 | 586.4691 |
| $\Delta_J$ (kHz) | 31.024 | 30.376 | 30.265 | 31.434 | 36.772 | 31.479 | 33.223 | 34.322 | 32.310 | 31.116 | 31.744 | 30.728 |
| $\Delta_K$ (kHz) | 20151.556 | 20008.733 | 19238.550 | 20091.078 | 31624.769 | 21159.931 | 21202.182 | 24202.347 | 20200.441 | 19520.362 | 20509.788 | 19913.362 |
| $\Delta_{JK}$ (kHz) | -608.438 | -597.478 | -564.093 | -615.926 | -801.235 | -617.467 | -661.344 | -631.358 | -623.002 | -617.499 | -626.043 | -617.799 |
| $\delta_J$ (kHz) | 4.191 | 4.087 | 4.024 | 4.262 | 4.839 | 4.300 | 4.588 | 4.555 | 4.371 | 4.223 | 4.296 | 4.135 |
| $\delta_K$ (kHz) | -48.252 | -19.301 | -105.699 | -25.123 | 93.813 | 36.675 | 62.739 | 96.278 | -93.955 | -14.335 | -14.462 | -36.573 |
| $|\mu_a|$ (D) | 0.55 | 0.62 | 0.72 | 0.52 | 0.32 | 0.43 | 0.45 | 0.28 | 0.47 | 0.62 | 0.50 | 0.59 |
| $|\mu_b|$ (D) | 0.20 | 0.29 | 0.31 | 0.19 | 0.15 | 0.14 | 0.20 | 0.22 | 0.10 | 0.33 | 0.18 | 0.32 |
| $|\mu_c|$ (D) | 1.22 | 1.24 | 1.26 | 1.23 | 1.02 | 1.16 | 1.19 | 1.15 | 1.20 | 1.27 | 1.22 | 1.26 |
| $|\mu_{TOT}|$ (D) | 1.35 | 1.41 | 1.49 | 1.34 | 1.08 | 1.24 | 1.29 | 1.20 | 1.29 | 1.45 | 1.33 | 1.43 |

Based on the optimized structures and microwave rotational constants $A$, $B$, and $C$, we further investigated whether Ac, Ac-w, and Ac-w$_2$ have rotational transitions. As we all know, a asymmetric molecule with $A>B>C$ interacts appreciably with a microwave electromagnetic field to emit or absorb radiation only if it has an electric or magnetic dipole moment[56], and a non-zero dipole moment component in $x$, $y$, or $z$ direction, which means that there can be corresponding $a$-, $b$- or $c$-type of transition. It is reported that revDSD-PBEP86-D3(BJ) is suitable to calculate the electron structures[35], so it can be good at calculating dipole moments. Due to the fact that the $|\mu_a|$, $|\mu_b|$, and $|\mu_{TOT}|$ calculated with revDSD-PBEP86-D3(BJ)/may-cc-pVTZ can basically match with the experimental data[9,55], so these data are mainly regarded as the prediction of dipole moment. For the SCFAs, they are all asymmetric-top molecules but the nearly symmetric planar, which causes the absence of $c$-dipole moment, indicating that the $c$-type transition cannot occur. Take AA as an example, after calculation, its $A$, $B$, $C$ constants are decreasing (11327.7036, 9492.0740, 5334.6625), so AA is an asymmetric-top molecule. and the $|\mu_a|$, $|\mu_b|$, and $|\mu_c|$ of it is in turn 0.86, 1.47, and 0.00 (**SI Table S1.1**), so it may have $a$- and $b$-type spectra and definitely cannot have a $c$-type spectrum. Like PTA-w, $|\mu_a|$, $|\mu_b|$, and $|\mu_c|$ selection rules are active, so three types of transitions may occur (the dipole moment data were summarized in **Table 2**). However, according to previous reports[9,57], tunnel effect can occur for the motions of uncoordinated H atoms, so some $c$-type spectra may not be observed in the clusters Ac-w and Ac-w$_2$.

3.2 Benchmark of rotational constants
To test the reliability of the selected twelve computational method, we compared the calculated data with the experimental data[8,9]. Due to the fact that the molecules mainly rotate in the ground state at low temperature and low pressure, the rotation constants *A*, *B* and *C* are enormously more important in the microwave rotational spectra, in which case we need to find the most suitable method for calculating *A*, *B* and *C*. As shown in **Fig. 1**, the absolute errors (AE) from the six experimental reference values[8,9] of *A* for AA, PPA and their corresponding hydrates and the correspondingly calculated data obtained by the twelve methods were set as the vertical axis. Moreover, we have performed linear regression for them to get the slope, intercept, and squared residual ($R^2$), mean absolute deviations (MADs) of slope and intercept, which were summarized in **Table S3** of **SI**.

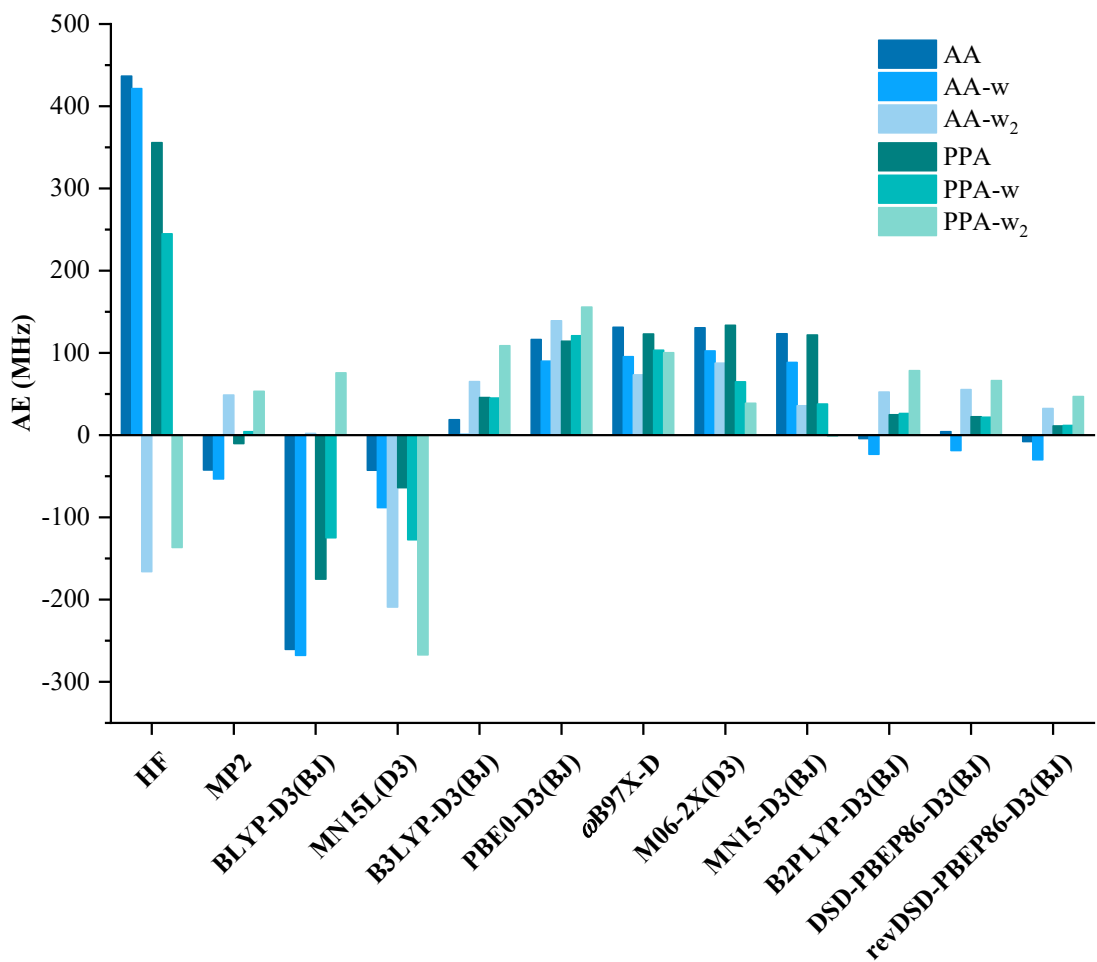

**Fig. 1** The AE for rotational constant *A* of AA, AA-w, AA-$w_2$, PPA, PPA-w, and PPA-$w_2$ by using twelve methods (may-cc-pVTZ basis set)

With the lower MADs and the closer slopes to 1, the DHDFs are the most suitable methods for the calculation of rotational constants for SCFAs and their hydrates, and the performance of revDSD-PBEP86-D3(BJ) with the slope of 0.99903 is the best one. Among the hybrid functionals, PBE0-D3(BJ), B3LYP-D3(BJ) and $\omega$B97X-D perform better than the other functionals. Furthermore, the linear regression result obtained by the HF method deviates the most. Although the MP2 method also has the good performance as the hybrid GGA functional, it has consumed relative long time.

The computational time of MP2 approximately equals to that of hybrid functional methods. For example, using HXA-$w_2$ as an example, the CPU running time for geometry optimization at the revDSD-PBEP86-D3(BJ)/may-cc-pVTZ level is 2 days and 20 hours, while the same basis set with MP2 method requires 3 days and 6 hours, making the MP2 method less cost-effective.

In addition, some studies advocate using the MP2 method for geometry optimization and then performing single-point correction to MP2 energy calculations using the CCSD(T) method (MP2+ΔCCSD(T)) to study interstellar molecular structures, molecular spectroscopic constants, and molecular energies[58]. Through the testing and discussion above, it was found that the revDSD-PBEP86-D3(BJ) method, which is comparable to the MP2 on structure optimization in terms of time, has advantages in optimizing structures and calculating rotational constants. Although the CCSD(T) method is more accurate than rev-DSD-PBEP86-D3(BJ) in calculating electronic energy, it cannot obtain information such as free energy correction quantities without the Hessian matrix for vibrational analysis in software such as *Gaussian*. Even if computable software is used, the required time is longer. This combined method is inferior to the rev-DSD-PBEP86-D3(BJ) method in terms of calculating electronic structure, electron density, dipole moment, rotational constants, thermodynamic energy, and computational time.

3.3 Stability of the clusters

Aimed at exploring whether the rotation of the Ac-w and Ac-$w_2$ in the vibrational ground state can still maintain the cluster morphology at low temperature and low pressure, we constructed four water-losing models: (i) the monohydrate loses one molecule of water, (ii) the water dimer loses two molecules of water, (iii) the water dimer loses one molecule of water, (iv) the water dimer loses another molecule of water. To study the strength and the composition of the hydrogen bonds, we did the SAPT energy decomposition calculation for the hydrated cluster, and decomposed the energy of hydrogen bond between molecules into electrostatic interactions, exchange mutual repulsion, induction, and dispersion. As can be seen from **Fig. 2**, the electrostatic attraction with positive contribution dominates in the hydrated clusters, whilst the dispersion and induction play an auxiliary role. The exact data for electrostatics, exchange, induction, dispersion and total SAPT2+(3) δMP2 energy is summarized in **Table S4**. In addition, according to the total SAPT2+(3)δMP2 in the figure, the decomposition of water dimer requires more energy to lose two water molecules, so the dimer is more stable. Furthermore, it can be known from **Table 2** and **Table S2.1-2.14** in **SI** that water restricts the rotation of the acid. That is to say, we can see that the more the coordination is corresponding to the smaller the rotational constant and the smaller the moment of inertia, so it is easier to rotate in the ground state, and the energy is lower. In conclusion, the SCFAs in the outer space will preferentially form more stable dihydrate clusters.

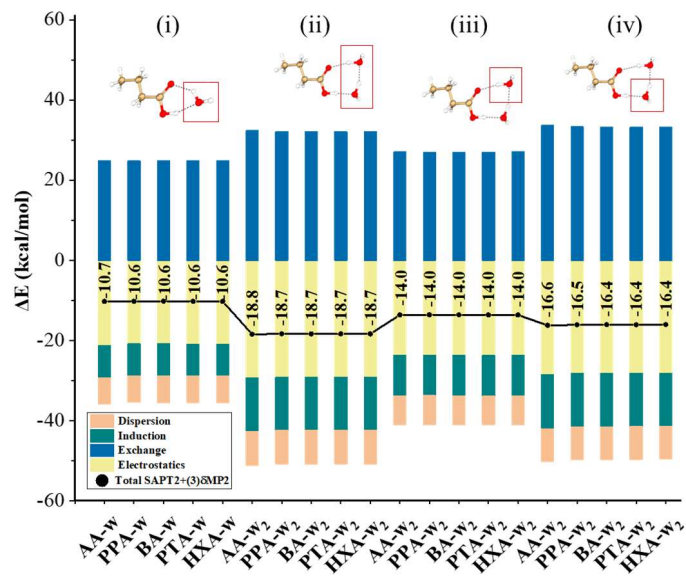

**Fig. 2** Decomposition energy diagram of hydrogen bonds in four types of clusters using SAPT

In order to explore whether such cluster structures can be maintained during rotation at different temperatures in the presence of water, both the stability of the two kinds of hydrates were investigated as follows. As we all know, SAPT only computes the electronic energy without the thermal corrections, here we further calculated the Gibbs free energies with the thermal corrections at different temperatures and the values are provided in **Table S5**. After that, $\Delta G$ of the dehydration procedure of water monomer and water dimer can be calculated (**Table S6**). It is found that the $\Delta G$ decreases with the rise of temperature so the stability declines.

Taking Gibbs-Helmholtz equation into the consideration, we set lg(reactant/product) as the ordinate, and its change with temperature is shown in **Fig. 3**. At 200 K and below, the ratio of the reactant (i.e. monohydrate or dihydrate cluster) to the product loses a molecule of water is greater than 1000:1, which indicates a great advantage in the form of clusters. In other words, rotation at higher temperatures than 200 K may lead to changes in the cluster skeleton.

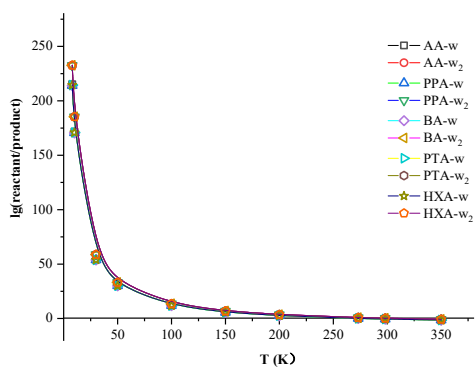

**Fig. 3** The relationship between the ratio of clusters and dehydrated molecules at different temperatures

3.4 Internal rotation

In order to explore whether the internal rotation of different acids occurs in the microwave rotation spectrum at low temperature and low pressure, we performed an energy surface scan on the $\Phi$ of the terminal methyl group to get $V_3$ barrier. The electron energy difference $\Delta E^{\neq}$ between the eclipsed and staggered conformation is shown in **Table 4**. We can see that $\Delta E^{\neq}$ is high, so when the rotor is regarded as a rigid rotor, the spectrum will not split without internal rotation at low temperature，theoretically. To explore the relationship between internal rotation and temperature, we located the rotational transition state (**SI**) and added the thermo corrections to the acids and clusters at 5 K, 50 K, 100 K and 298 K, which are shown in **Table 4** as well. Accompanied with the increasing temperature, the $\Delta G^{\neq}$ becomes larger, possibly enabling the methyl group to overcome the rotational energy barrier. It can be seen that the barriers for SCFAs and clusters containing three carbons and more are relatively high, we deduce that internal rotations for them cannot occur. But it cannot be confirmed that whether rotation of -$CH_3$ can occur in real situations, especially for AA and its clusters merely with the data of $\Delta E^{\neq}$ and $\Delta G^{\neq}$.

**Table 4** Internal rotational electronic energy barrier $\Delta E^{\neq}$ and the rotational Gibbs free energy barrier $\Delta G^{\neq}$ calculated at the level of revDSD-PBEP86-D3(BJ)/may-cc-pVTZ at different temperatures.

|  | $\Delta E^{\neq}$(kJ/mol) | $\Delta G^{\neq}$(5 K, kJ/mol) | $\Delta G^{\neq}$(50 K, kJ/mol) | $\Delta G^{\neq}$(100 K, kJ/mol) | $\Delta G^{\neq}$(298 K, kJ/mol) |
|---|---|---|---|---|---|
| **AA** | 1.64 | 1.51 | 1.58 | 1.92 | 4.74 |
| **AA-w** | 1.34 | 1.22 | 1.32 | 1.72 | 4.83 |
| **AA-w$_2$** | 1.12 | 1.03 | 1.16 | 1.63 | 5.02 |
| **PPA** | 10.10 | 9.35 | 9.37 | 9.48 | 10.91 |
| **PPA-w** | 10.22 | 9.41 | 9.43 | 9.53 | 10.88 |
| **PPA-w$_2$** | 10.35 | 9.56 | 9.58 | 9.68 | 11.03 |
| **BA** | 12.96 | 12.19 | 12.20 | 12.26 | 13.38 |
| **BA-w** | 12.98 | 12.19 | 12.20 | 12.25 | 13.31 |
| **BA-w$_2$** | 12.98 | 12.20 | 12.20 | 12.26 | 13.32 |
| **PTA** | 12.39 | 11.62 | 11.63 | 11.69 | 12.83 |
| **PTA-w** | 12.39 | 11.63 | 11.64 | 11.70 | 12.88 |
| **PTA-w$_2$** | 12.39 | 11.61 | 11.61 | 11.67 | 12.79 |
| **HXA** | 12.40 | 11.62 | 11.63 | 11.68 | 12.78 |
| **HXA-w** | 12.39 | 11.62 | 11.63 | 11.70 | 12.85 |
| **HXA-w$_2$** | 12.39 | 11.61 | 11.62 | 11.68 | 12.81 |

3.5 AIMD simulations

Considering the representativity and comparability, we selected the moderate SCFA monohydrates with even number of carbons (**Fig.4**) to study the dynamic behavior in real situations.

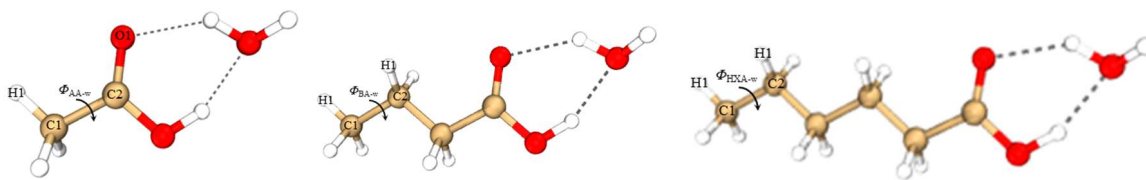

**Fig. 4** The structures and investigated dihedral angels of AA-w, BA-w and HXA-w

According to molecular dynamics we further verified whether the clusters are stable and whether internal rotations can occur at the temperatures of 5 K, 50 K, and 100 K. Atomic velocities were randomly initialized by Maxwell distribution AIMD simulations of 10 ps were performed at the PBE0(D3)/def2-TZVP level using a canonical sampling through velocity rescaling (CSVR) thermostat. The dihedral angle of terminal -$CH_3$ $\Phi$ is mainly investigated. Taking BA-w at 5 K as an example, we procured the trajectory based on AIMD, which is shown in **Fig. 4**. At this time scale, the dihedral angle

$\Phi_{BA-w}$ is twisted between 54.67° and 62.03° and the optimized dihedral angle is 58.20°, in which case, BA-w only performs small-scale thermal fluctuations in stable conformational space and there is no internal rotation at 5 K.

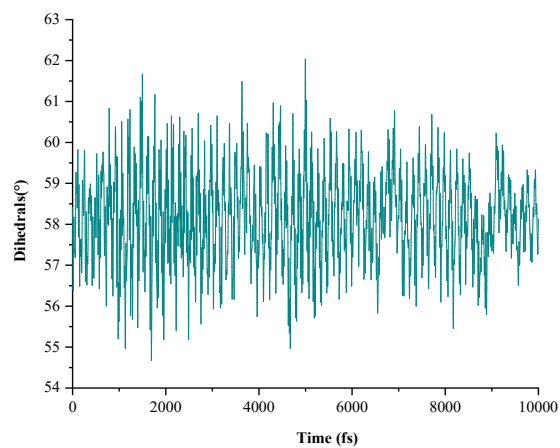

**Fig. 5** The change of $\Phi_{BA-w}$ in 10 ps (5 K)

At 5 K, we also provided the change of $\Phi_{AA-w}$ in **Fig. 6 (a)** and $\Phi_{HXA-w}$ in **SI** (**Fig. S3.1**). Both of the rotations of the methyl are small and cannot cause the spectral splitting. According to **Fig. 5**, **Fig. 6 (a)**, and **Fig. S3.1**, there are small differences for the dihedral angle, indicating the SCFAs can exist stably. Due to the fact that the experimental conditions of microwave spectroscopy are extremely low temperature (<10 K) in order to simulate the interstellar space, internal rotation cannot occur and no 3-fold splitting can be observed.

Then we explored the internal rotation of the methyl group with the rise of temperature. According to the optimized structure, the $\Phi_{AA-w}$ is -1.78°. At 5 K, it swings at a relatively small scale from -17.51° to 10.82° (**Fig. 6a**), it will not cross the internal rotation energy barrier. Nonetheless, internal rotation can occur at both 50 K and 100 K and AA-w rotate more freely at 100 K (**Fig. 6b, c**).

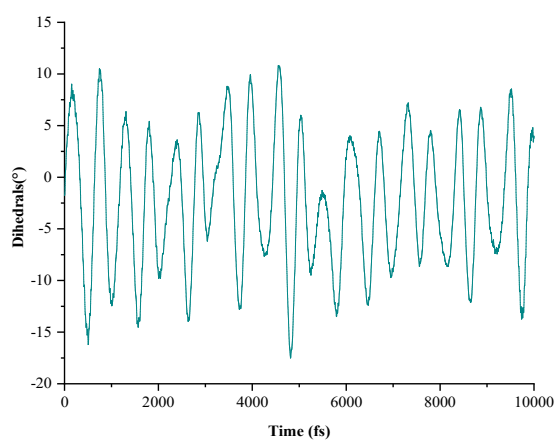

(a)

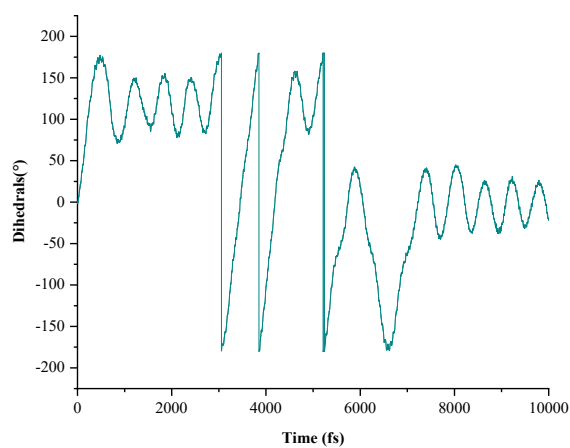

(b)

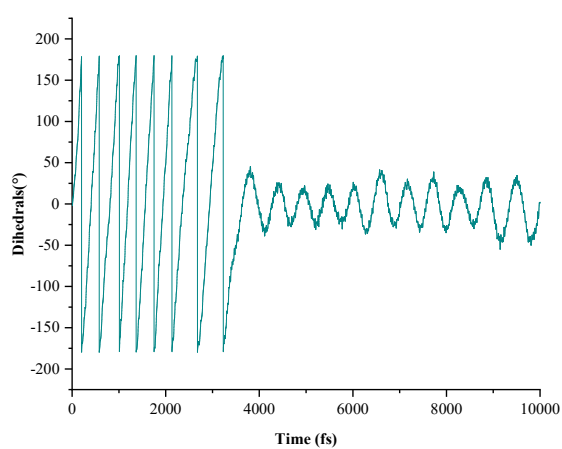

(c)

**Fig. 6 (a), (b), (c)** The change of $\Phi_{AA-w}$ in 10 ps at 5 K, 50 K and 100 K

Applied with SPCAT, we used the experimental data[9] to predict the microwave rotational spectroscopy. The spectral line files of the AA-w molecules were visualized with pgopher and shown in **Fig. 7 (a), (b), (c)**. It can be seen that the spectra unsplit at 5 K, and split at 50 K and 100 K. The corresponding methyl group does not rotate at 5 K, but rotates at 50 K and 100 K, and the figure is consistent with the results obtained by AIMD (**Fig. 6 (a), (b), (c)**).

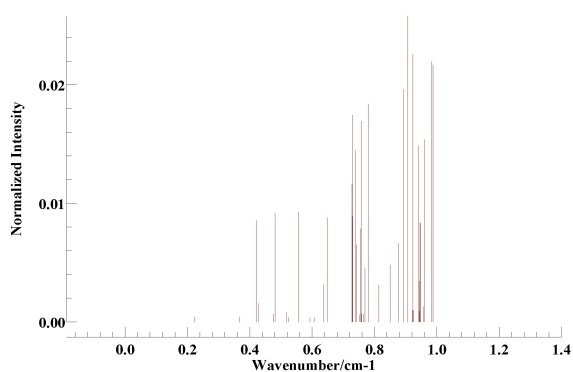

(a)

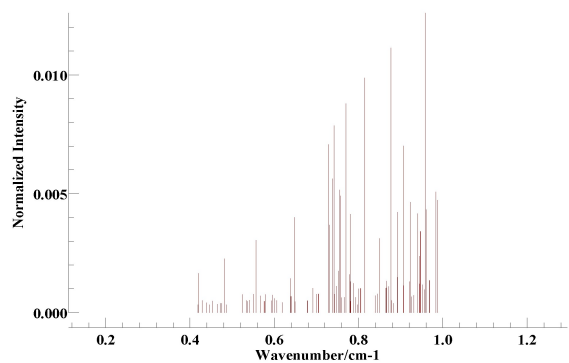

(b)

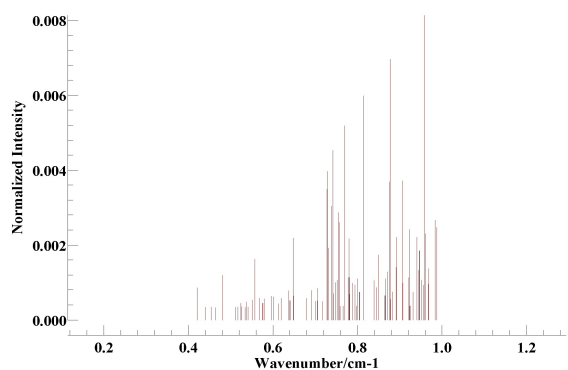

(c)

**Fig. 7 (a), (b), (c)** Microwave rotational spectra of AA-w at 5 K, 50 K and 100 K

However, in **Fig. S3.2-3.5**, we provided the dihedral rotation angle of the 10 ps AIMD simulation of BA-w and HXA-w at 50 K and 100 K, and the change of $\Phi$ were both small and no internal rotation occurred, which can perfectly correspond to the conclusion based on energy barrier.

In order to investigate why only AA-w can rotate at high temperature, we further compared the structure of AA-w and BA-w for the difference of the rotation of $\Phi$ in **Fig.8**. For the van der Waals radius of H, O atoms are 1.0 Å and 1.35 Å, the minimum distance that atoms in can interact with each other are $r_{min}(H…H)=2.0$ Å and $r_{min}(H…O)=2.5$ Å, respectively[59]. For AA-w, the distance $r(H…O)$ of H1 and O1 is 2.6 Å when $\Phi$ equals to zero, so the methyl can rotate with sufficient energy provided by temperature. However, $r(H…H)$ of BA-w is 2.3 Å, lower than 2.5 Å, which means van der Waals interaction exists.

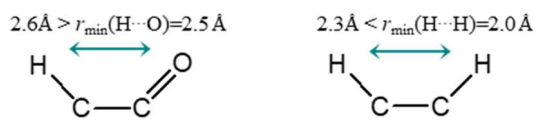

**Fig.8** The contracs of the intramolecular interaction of H atom from terminal -CH3 and O or H atom on adjacent carbon for AA-w and BA-w

It can be concluded that the presence of the methylene group hinders the rotation of the terminal methyl group.

# 4. Conclusion

The geometries of SCFAs have been optimized using twelve computational methods, respectively, which provide rotational spectroscopic parameters (rotational and centrifugal distortion constants) of target molecules, directly reflected by their mass distribution. Taken experimental values of rotational spectroscopic parameters by high resolution rotational spectroscopy as the benchmark, the revDSD-PBEP86-D3(BJ) was determined to be the optimal calculation method, and the DHDFs have excellent performance. When the computing resources are limited, PBE0, B3LYP, and $\omega$B97X-D can be selected for ordinary functionals to calculate the rotational spectra of such substances.

The decomposing energy of SCFA…water bonds in the hydrated clusters indicated that the electrostatic attraction was dominant in the hydrated clusters, and there were also dispersion and dissociation effects. By analyzing rotation constants, it is known that SCFAs in space will preferentially form more stable dihydrate clusters. In addition, by penetrating into the water loss reaction process, it is known that the critical temperature for maintaining the hydrated cluster structure is approximately equal to 200 K.

Through analyzing $\Delta G^{\neq}$ and $\Delta E^{\neq}$ from potential energy surface scanning and transition state optimization, the possibility of internal rotation can be conjecture. According to AIMD simulations, the real dynamic change of the dihedral angle of the terminal methyl group provides theoretical support for whether three-fold splitting occurs due to internal rotation, and innovatively opens up new ideas for predicting splitting. In conclusion, SCFAs exist stably at 5 K to 100 K and $\Phi$s of SCFAs do not undergo internal rotation at 5 K. $\Phi_{AA-w}$ rotates at 50 K and above, but for SCFAs with three or more carbons, the methyl groups are difficult to rotate even at 100 K. We verified the existence of methylene group hinders the rotation of the terminal methyl group.

In addition, the parameter data that have not been reported experimentally can be used as the theoretical prediction result, because the computational accuracy was tested to be high enough.

wrote them. in *Advances in Protein Chemistry* (eds. Anfinsen, C. B., Anson, M. L., Edsall, J. T. & Richards, F. M.) vol. 23 283–437 (Academic Press, 1968).

## Supporting Information

### 1. Statistical measures of the data set

For a set of calculated data $\{a_1, a_2, \ldots, a_n\}$ and the experimental reference data $\{b_1, b_2, \ldots, b_n\}$ statistical meaures are:

(1) Root mean square deviation (RMSD):

$$RMSD = \sqrt{\frac{1}{n}\sum_{i=1}^{n}|a_i - b_i|^2}$$

(2) Absolute deviation (ADEV):

$$AE = a_i - b_i$$

(3) Mean absolute deviation (MAD):

$$MAD = \frac{1}{n}\sum_{i=1}^{n}|a_i - b_i|$$

### 2. RMSD of optimized geometries

We used RMSD to measure geometric differences to check whether structures optimized with these methods are plausible. Taking the optimal rev-DSDPBEP86-D3(BJ) as the benchmark, the RMSDs of the coordinates of the optimized geometries calculated with different methods were summarized in **Table 3**. Despite being mediocre in calculating the rotational constants of SCFAs, MP2 and $\omega$B97X-D are outstanding in calculating geometric structures. The rest of the ordinary functionals are not much different except for MN15L (D3), and HF has the worst performance.

**Table S1** RMSDs of SCFAs geometries calculated with different methods relative to revDSD-PBEP86-D3(BJ)

| RMSD(Å) | B2PLYP-D3(BJ) | B3LYP-D3(BJ) | BLYP-D3(BJ) | DSD-PBEP86-D3(BJ) | HF | M06-2X(D3) | MN15-D3(BJ) | MN15L(D3) | MP2 | PBE0-D3(BJ) | $\omega$B97X-D |
|---|---|---|---|---|---|---|---|---|---|---|---|
| AA | 0.016 | 0.005 | 0.005 | 0.008 | 0.004 | 0.005 | 0.002 | 0.004 | 0.016 | 0.001 | 0.005 |
| AA-w | 0.006 | 0.012 | 0.025 | 0.008 | 0.108 | 0.020 | 0.011 | 0.074 | 0.017 | 0.037 | 0.017 |
| AA-w$_2$ | 0.008 | 0.126 | 0.025 | 0.030 | 0.102 | 0.029 | 0.043 | 0.009 | 0.020 | 0.036 | 0.023 |
| PPA | 0.002 | 0.018 | 0.008 | 0.012 | 0.013 | 0.006 | 0.005 | 0.004 | 0.009 | 0.022 | 0.007 |
| PPA-w | 0.103 | 0.019 | 0.017 | 0.068 | 0.012 | 0.034 | 0.006 | 0.011 | 0.026 | 0.008 | 0.017 |
| PPA-w$_2$ | 0.009 | 0.021 | 0.041 | 0.008 | 0.116 | 0.029 | 0.031 | 0.095 | 0.012 | 0.041 | 0.022 |
| BA | 0.003 | 0.007 | 0.022 | 0.002 | 0.015 | 0.007 | 0.011 | 0.012 | 0.007 | 0.006 | 0.005 |
| BA-w | 0.005 | 0.010 | 0.026 | 0.008 | 0.094 | 0.018 | 0.013 | 0.065 | 0.012 | 0.033 | 0.015 |
| BA-w$_2$ | 0.009 | 0.019 | 0.041 | 0.008 | 0.109 | 0.030 | 0.029 | 0.089 | 0.013 | 0.039 | 0.021 |
| PTA | 0.004 | 0.009 | 0.026 | 0.003 | 0.017 | 0.008 | 0.014 | 0.014 | 0.008 | 0.007 | 0.006 |
| PTA-w | 0.006 | 0.011 | 0.029 | 0.007 | 0.090 | 0.018 | 0.018 | 0.061 | 0.013 | 0.031 | 0.015 |
| PTA-w$_2$ | 0.010 | 0.021 | 0.046 | 0.008 | 0.103 | 0.035 | 0.035 | 0.089 | 0.015 | 0.037 | 0.022 |
| HXA | 0.004 | 0.008 | 0.028 | 0.003 | 0.015 | 0.008 | 0.013 | 0.014 | 0.008 | 0.008 | 0.005 |
| HXA-w | 0.005 | 0.010 | 0.031 | 0.007 | 0.084 | 0.017 | 0.016 | 0.059 | 0.013 | 0.030 | 0.014 |

| | | | | | | | | | | | | |
|---|---|---|---|---|---|---|---|---|---|---|---|---|
| HXA-w$_2$ | 0.009 | 0.020 | 0.047 | 0.008 | 0.097 | 0.035 | 0.034 | 0.085 | 0.015 | 0.036 | 0.021 | |
| Average | 0.013 | 0.021 | 0.028 | 0.013 | 0.065 | 0.020 | 0.019 | 0.046 | 0.013 | 0.025 | 0.014 | |

## 3. Rotational constants, centrifugal distortion constants and dipole moment of SCFAs and corresponding water clusters (except for PTA-w) calculated with twelve different methods

**Table S2.1** Rotational constants, centrifugal distortion constants and dipole moment of AA calculated with twelve different methods

| Parameters | Experiment | B2PLYP-D3(BJ) | B3LYP-D3(BJ) | BLYP-D3(BJ) | DSD-PBEP86-D3(BJ) | HF | M06-2X(D3) | MN15-D3(BJ) | MN15L(D3) | MP2 | PBE0-D3(BJ) | revDSD-PBEP86-D3(BJ) | ωB97X-D |
|---|---|---|---|---|---|---|---|---|---|---|---|---|---|
| $A$ (MHz) | 11335.47 | 11331.3380 | 11354.3677 | 11074.9837 | 11339.7891 | 11772.1531 | 11466.0607 | 11458.9316 | 11292.6789 | 11293.0493 | 11451.7509 | 11327.7036 | 11466.6232 |
| $B$ (MHz) | 9478.604 | 9499.3886 | 9481.9171 | 9329.9409 | 9507.2454 | 9602.8684 | 9531.7030 | 9512.7027 | 9425.5035 | 9530.1354 | 9558.2884 | 9492.0740 | 9523.8773 |
| $C$ (MHz) | 5324.906 | 5336.6727 | 5336.5581 | 5228.5165 | 5341.9304 | 5464.8743 | 5377.1375 | 5369.2065 | 5307.7079 | 5337.8268 | 5382.4041 | 5334.6625 | 5374.7139 |
| $\Delta_J$ (kHz) | N/A | 9446.064 | 9481.687 | 9776.197 | 9337.806 | 8655.907 | 9212.378 | 9571.894 | 9013.473 | 9340.936 | 9375.644 | 9339.099 | 9312.217 |
| $\Delta_K$ (kHz) | N/A | -8035.385 | -3533.547 | -3978.007 | -4108.203 | -662.922 | -2731.057 | -1632.531 | -3371.541 | -20423.482 | -3541.590 | -4417.788 | -3684.501 |
| $\Delta_{JK}$ (kHz) | N/A | -569.356 | -5100.388 | -4916.918 | -4400.892 | -7206.578 | -5664.919 | -7122.248 | -4846.207 | 11907.662 | -5008.746 | -4090.709 | -4796.019 |
| $\delta_J$ (kHz) | N/A | 676.434 | 665.809 | 687.381 | 667.086 | 602.456 | 636.306 | 710.378 | 624.922 | 666.783 | 661.747 | 664.119 | 661.007 |
| $\delta_K$ (kHz) | N/A | 4035.855 | 5612.186 | 5316.169 | 6496.997 | 5959.211 | 6650.158 | 8207.777 | 8702.987 | -1045.800 | 6562.557 | 6288.031 | 5777.826 |
| $|\mu_a|$ (D) | 0.76 | 0.79 | 0.88 | 0.89 | 0.77 | 0.91 | 0.89 | 0.87 | 0.89 | 0.75 | 0.87 | 0.76 | 0.87 |
| $|\mu_b|$ (D) | 1.46 | 1.54 | 1.55 | 1.48 | 1.53 | 1.74 | 1.55 | 1.58 | 1.57 | 1.45 | 1.58 | 1.52 | 1.59 |
| $|\mu_c|$ (D) | 0.00 | 0.00 | 0.00 | 0.00 | 0.00 | 0.00 | 0.00 | 0.00 | 0.00 | 0.00 | 0.00 | 0.00 | 0.00 |
| $|\mu_{TOT}|$ (D) | 1.7 | 1.73 | 1.78 | 1.72 | 1.71 | 1.96 | 1.79 | 1.80 | 1.81 | 1.63 | 1.80 | 1.70 | 1.81 |

**Table S2.2** Rotational constants, centrifugal distortion constants and dipole moment of AA-w calculated with twelve different methods

| Parameters | Experiment | B2PLYP-D3(BJ) | B3LYP-D3(BJ) | BLYP-D3(BJ) | DSD-PBEP86-D3(BJ) | HF | M06-2X(D3) | MN15-D3(BJ) | MN15L(D3) | MP2 | PBE0-D3(BJ) | revDSD-PBEP86-D3(BJ) | ωB97X-D |
|---|---|---|---|---|---|---|---|---|---|---|---|---|---|
| $A$ (MHz) | 11059.8286 | 11036.3939 | 11060.6092 | 10792.0194 | 11041.1119 | 11481.2759 | 11162.3810 | 11148.3107 | 10971.7582 | 11006.3978 | 11149.7825 | 11029.8567 | 11155.2643 |
| $B$ (MHz) | 2584.53109 | 2645.7208 | 2646.3109 | 2617.5750 | 2650.5678 | 2453.8643 | 2635.6970 | 2632.4780 | 2474.2669 | 2655.5473 | 2700.2312 | 2631.6705 | 2631.4093 |
| $C$ (MHz) | 2127.47530 | 2171.1810 | 2172.9801 | 2144.6421 | 2174.9874 | 2051.9078 | 2168.0917 | 2165.9459 | 2052.1173 | 2176.0837 | 2213.4167 | 2161.8049 | 2166.8892 |
| $\Delta_J$ (kHz) | 1.29 | 1183.167 | 1157.570 | 1161.696 | 1190.441 | 1587.217 | 1212.459 | 1235.178 | 1407.634 | 1243.645 | 1170.474 | 1217.337 | 1201.698 |
| $\Delta_K$ (kHz) | N/A | 5372.826 | 5291.078 | 5553.558 | 5352.061 | 1988.127 | 5122.312 | 5755.144 | 4897.632 | 5471.302 | 5408.876 | 5228.264 | 4807.190 |
| $\Delta_{JK}$ (kHz) | 3.38 | 2392.629 | 2403.949 | 2269.261 | 2318.571 | 5377.145 | 2570.245 | 2076.996 | 2820.726 | 2299.680 | 2085.443 | 2444.342 | 2763.427 |
| $\delta_J$ (kHz) | 0.271 | 226.055 | 220.593 | 223.119 | 227.492 | 275.935 | 230.079 | 234.522 | 256.507 | 238.149 | 224.598 | 231.364 | 226.793 |
| $\delta_K$ (kHz) | N/A | 1298.764 | 1626.398 | 727.025 | 1350.052 | 5386.752 | 2809.820 | 2631.054 | 3292.875 | 597.491 | 1303.812 | 1677.164 | 2115.216 |
| $|\mu_a|$ (D) | 0.2-0.3 | 0.44 | 0.52 | 0.61 | 0.42 | 0.19 | 0.32 | 0.39 | 0.26 | 0.33 | 0.55 | 0.41 | 0.52 |
| $|\mu_b|$ (D) | 0.2 | 0.25 | 0.31 | 0.31 | 0.24 | 0.27 | 0.23 | 0.27 | 0.32 | 0.15 | 0.35 | 0.23 | 0.34 |
| $|\mu_c|$ (D) | N/A | 1.22 | 1.24 | 1.27 | 1.23 | 1.02 | 1.16 | 1.20 | 1.15 | 1.18 | 1.27 | 1.22 | 1.26 |
| $|\mu_{TOT}|$ (D) | N/A | 1.32 | 1.38 | 1.44 | 1.32 | 1.07 | 1.23 | 1.29 | 1.22 | 1.23 | 1.43 | 1.31 | 1.41 |

**Table S2.3** Rotational constants, centrifugal distortion constants and dipole moment of AA-w$_2$ calculated with twelve different methods

| Parameters | Experiment | B2PLYP-D3(BJ) | B3LYP-D3(BJ) | BLYP-D3(BJ) | DSD-PBEP86-D3(BJ) | HF | M06-2X(D3) | MN15-D3(BJ) | MN15L(D3) | MP2 | PBE0-D3(BJ) | revDSD-PBEP86-D3(BJ) | ωB97X-D |
|---|---|---|---|---|---|---|---|---|---|---|---|---|---|
| $A$ (MHz) | 4445.466 | 4498.0012 | 4510.7249 | 4447.5348 | 4500.8261 | 4279.3482 | 4532.9982 | 4481.1428 | 4236.2670 | 4494.3074 | 4584.6877 | 4477.9266 | 4518.6978 |
| $B$ (MHz) | 1618.43953 | 1645.7909 | 1654.0597 | 1636.4068 | 1646.1454 | 1546.1384 | 1651.1492 | 1627.3078 | 1531.0366 | 1645.1127 | 1681.1677 | 1636.7481 | 1649.4601 |
| $C$ (MHz) | 1199.67784 | 1218.8660 | 1224.6847 | 1211.2463 | 1219.4890 | 1145.8735 | 1224.3032 | 1207.0002 | 1137.4458 | 1218.2618 | 1245.4942 | 1212.6185 | 1223.2403 |

| Parameters | Experiment | B2PLYP-D3(BJ) | B3LYP-D3(BJ) | BLYP-D3(BJ) | DSD-PBEP86-D3(BJ) | HF | M06-2X(D3) | MN15-D3(BJ) | MN15L(D3) | MP2 | PBE0-D3(BJ) | revDSD-PBEP86-D3(BJ) | ωB97X-D |
|---|---|---|---|---|---|---|---|---|---|---|---|---|---|
| $\Delta_J$ (kHz) | 0.364 | 289.324 | 279.269 | 278.232 | 294.530 | 392.075 | 318.178 | 311.055 | 376.118 | 309.538 | 286.625 | 299.265 | 286.268 |
| $\Delta_K$ (kHz) | N/A | 542.263 | 530.721 | 503.826 | 533.401 | 1861.030 | 530.545 | 306.057 | 752.670 | 603.630 | 502.546 | 614.845 | 624.375 |
| $\Delta_{JK}$ (kHz) | 4.20 | 3268.063 | 3136.715 | 2975.276 | 3349.313 | 4354.608 | 3881.799 | 3983.836 | 5047.041 | 3451.922 | 3137.106 | 3368.858 | 3176.649 |
| $\delta_J$ (kHz) | 0.0788 | 64.319 | 61.529 | 60.720 | 65.111 | 90.846 | 68.353 | 69.366 | 81.649 | 67.870 | 62.411 | 66.328 | 62.716 |
| $\delta_K$ (kHz) | 2.76 | 1668.891 | 1634.757 | 1473.736 | 1722.752 | 2514.558 | 2098.945 | 2200.897 | 2842.248 | 1714.998 | 1626.777 | 1767.259 | 1724.531 |
| $|\mu_a|$ (D) | 0.2-0.3 | 0.25 | 0.32 | 0.36 | 0.24 | 0.27 | 0.20 | 0.19 | 0.32 | 0.16 | 0.30 | 0.24 | 0.38 |
| $|\mu_b|$ (D) | <0.1 | 0.03 | 0.07 | 0.07 | 0.01 | 0.04 | 0.03 | 0.04 | 0.00 | 0.06 | 0.10 | 0.00 | 0.07 |
| $|\mu_c|$ (D) | N/A | 2.18 | 2.23 | 2.31 | 2.20 | 1.66 | 2.13 | 2.08 | 2.10 | 2.14 | 2.29 | 2.19 | 2.26 |
| $|\mu_{TOT}|$ (D) | N/A | 2.20 | 2.25 | 2.34 | 2.22 | 1.68 | 2.14 | 2.09 | 2.12 | 2.15 | 2.31 | 2.21 | 2.29 |

**Table S2.4** Rotational constants, centrifugal distortion constants and dipole moment of PPA calculated with twelve different methods

| Parameters | Experiment | B2PLYP-D3(BJ) | B3LYP-D3(BJ) | BLYP-D3(BJ) | DSD-PBEP86-D3(BJ) | HF | M06-2X(D3) | MN15-D3(BJ) | MN15L(D3) | MP2 | PBE0-D3(BJ) | revDSD-PBEP86-D3(BJ) | ωB97X-D |
|---|---|---|---|---|---|---|---|---|---|---|---|---|---|
| $A$ (MHz) | 10,155.38 | 10180.1036 | 10201.2675 | 9980.3165 | 10177.8194 | 10510.9915 | 10289.1544 | 10277.2007 | 10091.5021 | 10145.1818 | 10269.7348 | 10166.6671 | 10278.2855 |
| $B$ (MHz) | 3,817.89 | 3823.6489 | 3809.1360 | 3742.3173 | 3837.1509 | 3842.0012 | 3851.3407 | 3859.4848 | 3818.5616 | 3850.4956 | 3849.9794 | 3828.7849 | 3829.6043 |
| $C$ (MHz) | 2,875.17 | 2876.8315 | 2870.4162 | 2816.1146 | 2884.7982 | 2912.3757 | 2901.4243 | 2904.9340 | 2868.7878 | 2889.2621 | 2899.1084 | 2879.3153 | 2888.1554 |
| $\Delta_J$ (kHz) | 0.685 | 754.751 | 751.864 | 761.116 | 756.633 | 692.510 | 749.130 | 779.128 | 729.262 | 761.787 | 768.495 | 755.471 | 746.535 |
| $\Delta_K$ (kHz) | 2.89 | 3026.239 | 3107.928 | 3087.101 | 3030.146 | 2979.709 | 3301.862 | 3501.249 | 3188.791 | 2938.086 | 3012.953 | 3037.365 | 3092.001 |
| $\Delta_{JK}$ (kHz) | 3.78 | 3548.039 | 3479.770 | 3680.852 | 3475.076 | 3031.833 | 3142.119 | 3275.279 | 3112.626 | 3564.123 | 3516.084 | 3463.778 | 3393.546 |
| $\delta_J$ (kHz) | -0.1866 | 188.105 | 186.851 | 189.174 | 189.149 | 169.804 | 187.462 | 195.773 | 183.326 | 191.217 | 191.728 | 188.713 | 185.661 |
| $\delta_K$ (kHz) | -0.0076 | -226.002 | 65.034 | -472.524 | 11.586 | 554.702 | 829.313 | 956.775 | 977.410 | -359.582 | 162.774 | 55.322 | 42.653 |
| $|\mu_a|$ (D) | 0.19 | 0.18 | 0.15 | 0.23 | 0.14 | 0.16 | 0.16 | 0.09 | 0.04 | 0.16 | 0.13 | 0.15 | 0.13 |
| $|\mu_b|$ (D) | 1.53 | 1.59 | 1.56 | 1.58 | 1.56 | 1.82 | 1.64 | 1.65 | 1.66 | 1.49 | 1.65 | 1.56 | 1.67 |
| $|\mu_c|$ (D) | 0.00 | 0.00 | 0.00 | 0.00 | 0.00 | 0.00 | 0.00 | 0.00 | 0.00 | 0.00 | 0.00 | 0.00 | 0.00 |
| $|\mu_{TOT}|$ (D) | 1.54 | 1.60 | 1.57 | 1.60 | 1.57 | 1.82 | 1.64 | 1.65 | 1.66 | 1.50 | 1.66 | 1.56 | 1.68 |

**Table S2.5** Rotational constants, centrifugal distortion constants and dipole moment of PPA-w calculated with twelve different methods

| Parameters | Experiment | B2PLYP-D3(BJ) | B3LYP-D3(BJ) | BLYP-D3(BJ) | DSD-PBEP86-D3(BJ) | HF | M06-2X(D3) | MN15-D3(BJ) | MN15L(D3) | MP2 | PBE0-D3(BJ) | revDSD-PBEP86-D3(BJ) | ωB97X-D |
|---|---|---|---|---|---|---|---|---|---|---|---|---|---|
| $A$ (MHz) | 8,167.31 | 8193.7784 | 8212.6059 | 8042.3866 | 8189.0616 | 8412.1626 | 8232.1756 | 8205.1112 | 8039.9723 | 8171.4096 | 8288.2501 | 8179.0910 | 8270.5710 |
| $B$ (MHz) | 1,618.81 | 1646.7225 | 1644.0013 | 1622.5808 | 1651.9176 | 1554.2992 | 1649.2113 | 1651.3609 | 1571.5648 | 1656.5415 | 1673.3163 | 1641.8665 | 1639.1736 |
| $C$ (MHz) | 1,377.16 | 1398.0436 | 1396.8327 | 1377.1568 | 1401.8540 | 1335.1135 | 1400.4938 | 1401.4404 | 1339.5501 | 1404.4276 | 1420.4550 | 1394.3366 | 1395.1658 |
| $\Delta_J$ (kHz) | 0.4115 | 349.824 | 341.485 | 338.988 | 353.318 | 463.522 | 350.753 | 368.019 | 406.661 | 367.444 | 347.290 | 360.593 | 350.884 |
| $\Delta_K$ (kHz) | N/A | 13066.437 | 13048.694 | 12439.221 | 12877.369 | 23656.485 | 13440.350 | 13090.659 | 15685.596 | 12900.575 | 12302.475 | 13331.862 | 13304.101 |
| $\Delta_{JK}$ (kHz) | -0.903 | -621.822 | -608.119 | -514.784 | -642.364 | -883.905 | -439.863 | -623.294 | -364.161 | -660.643 | -682.205 | -662.582 | -764.830 |
| $\delta_J$ (kHz) | -0.0757 | 64.166 | 62.496 | 61.725 | 64.939 | 81.014 | 63.497 | 67.327 | 71.051 | 67.501 | 63.914 | 66.068 | 64.382 |
| $\delta_K$ (kHz) | -0.0071 | 606.272 | 644.681 | 557.048 | 624.443 | 1330.522 | 799.604 | 841.196 | 1101.774 | 529.817 | 625.292 | 664.579 | 666.157 |
| $|\mu_a|$ (D) | N/A | 0.45 | 0.54 | 0.63 | 0.43 | 0.23 | 0.34 | 0.37 | 0.22 | 0.36 | 0.54 | 0.41 | 0.52 |
| $|\mu_b|$ (D) | N/A | 0.19 | 0.26 | 0.28 | 0.18 | 0.16 | 0.14 | 0.19 | 0.23 | 0.09 | 0.31 | 0.16 | 0.30 |
| $|\mu_c|$ (D) | N/A | 1.22 | 1.24 | 1.27 | 1.23 | 1.02 | 1.16 | 1.19 | 1.15 | 1.21 | 1.27 | 1.23 | 1.26 |
| $|\mu_{TOT}|$ (D) | N/A | 1.32 | 1.38 | 1.45 | 1.31 | 1.06 | 1.22 | 1.26 | 1.20 | 1.26 | 1.42 | 1.30 | 1.40 |

**Table S2.6** Rotational constants, centrifugal distortion constants and dipole moment of PPA-w$_2$ calculated with twelve different methods

| Parameters | Experiment | B2PLYP-D3(BJ) | B3LYP-D3(BJ) | BLYP-D3(BJ) | DSD-PBEP86-D3(BJ) | HF | M06-2X(D3) | MN15-D3(BJ) | MN15L(D3) | MP2 | PBE0-D3(BJ) | revDSD-PBEP86-D3(BJ) | ωB97X-D |
|---|---|---|---|---|---|---|---|---|---|---|---|---|---|
| $A$ (MHz) | 3,627.72 | 3706.0613 | 3736.4046 | 3703.4859 | 3694.2053 | 3491.0056 | 3666.4130 | 3626.5857 | 3360.4212 | 3681.0892 | 3783.5229 | 3674.6566 | 3727.9567 |
| $B$ (MHz) | 1,102.65 | 1112.1951 | 1112.2212 | 1095.6428 | 1115.5885 | 1064.8392 | 1127.8289 | 1116.5197 | 1074.6587 | 1117.6333 | 1130.3121 | 1109.7073 | 1112.8593 |
| $C$ (MHz) | 857.26038 | 867.0414 | 868.8696 | 857.5043 | 868.5882 | 825.2324 | 874.2157 | 865.1006 | 825.1862 | 868.9049 | 882.6898 | 863.9255 | 869.0290 |
| $\Delta_J$ (kHz) | 0.145 | 130.321 | 124.720 | 120.770 | 133.880 | 181.119 | 149.680 | 153.970 | 193.026 | 138.636 | 126.851 | 135.963 | 129.391 |
| $\Delta_K$ (kHz) | N/A | 3534.372 | 3416.831 | 3213.924 | 3591.334 | 5280.822 | 4224.068 | 4246.654 | 5682.303 | 3641.721 | 3374.421 | 3652.316 | 3604.883 |
| $\Delta_{JK}$ (kHz) | 0.692 | 541.365 | 515.994 | 528.617 | 554.717 | 662.033 | 601.667 | 545.434 | 599.917 | 599.739 | 536.877 | 560.144 | 514.953 |
| $\delta_J$ (kHz) | -0.0415 | 31.448 | 29.750 | 27.900 | 32.410 | 46.435 | 37.203 | 39.623 | 52.965 | 33.436 | 29.876 | 32.948 | 31.062 |
| $\delta_K$ (kHz) | -0.00841 | 554.938 | 529.363 | 493.861 | 575.351 | 817.054 | 687.033 | 703.488 | 916.871 | 585.756 | 536.232 | 583.587 | 561.062 |
| $|\mu_a|$ (D) | N/A | 0.29 | 0.36 | 0.42 | 0.28 | 0.32 | 0.25 | 0.21 | 0.33 | 0.22 | 0.34 | 0.29 | 0.42 |
| $|\mu_b|$ (D) | N/A | 0.07 | 0.02 | 0.00 | 0.08 | 0.04 | 0.13 | 0.08 | 0.04 | 0.16 | 0.02 | 0.09 | 0.01 |
| $|\mu_c|$ (D) | N/A | 2.18 | 2.23 | 2.32 | 2.20 | 1.66 | 2.14 | 2.09 | 2.12 | 2.16 | 2.29 | 2.19 | 2.27 |
| $|\mu_{TOT}|$ (D) | N/A | 2.20 | 2.26 | 2.35 | 2.22 | 1.69 | 2.16 | 2.10 | 2.14 | 2.18 | 2.32 | 2.21 | 2.31 |

**Table S2.7** Rotational constants, centrifugal distortion constants and dipole moment of BA calculated with twelve different methods

| Parameters | B2PLYP-D3(BJ) | B3LYP-D3(BJ) | BLYP-D3(BJ) | DSD-PBEP86-D3(BJ) | HF | M06-2X(D3) | MN15-D3(BJ) | MN15L(D3) | MP2 | PBE0-D3(BJ) | revDSD-PBEP86-D3(BJ) | ωB97X-D |
|---|---|---|---|---|---|---|---|---|---|---|---|---|
| $A$ (MHz) | 9022.7857 | 9040.2067 | 8843.9453 | 9021.4938 | 9304.9207 | 9116.0935 | 9110.4725 | 8943.4974 | 8991.8901 | 9106.7159 | 9011.3326 | 9107.1481 |
| $B$ (MHz) | 1837.3317 | 1830.9973 | 1801.1479 | 1842.7817 | 1842.2081 | 1848.0618 | 1853.0546 | 1831.3333 | 1849.1586 | 1848.7194 | 1838.9146 | 1839.4005 |
| $C$ (MHz) | 1570.2262 | 1566.1589 | 1538.9444 | 1574.3951 | 1581.7415 | 1580.9749 | 1584.4086 | 1564.3605 | 1577.9612 | 1581.1601 | 1571.3229 | 1574.3373 |
| $\Delta_J$ (kHz) | 142.856 | 142.646 | 144.891 | 142.982 | 130.316 | 141.823 | 147.962 | 137.969 | 144.059 | 144.698 | 142.760 | 141.092 |
| $\Delta_K$ (kHz) | 4381.479 | 4404.951 | 4482.283 | 4334.067 | 4056.244 | 4406.452 | 4681.089 | 4245.727 | 4325.496 | 4369.992 | 4330.698 | 4307.297 |
| $\Delta_{JK}$ (kHz) | 1452.466 | 1434.370 | 1492.573 | 1457.810 | 1254.524 | 1381.337 | 1440.858 | 1328.603 | 1488.604 | 1482.796 | 1451.737 | 1421.521 |
| $\delta_J$ (kHz) | 21.822 | 21.708 | 22.149 | 21.887 | 19.357 | 21.637 | 22.778 | 21.301 | 22.209 | 22.041 | 21.830 | 21.422 |
| $\delta_K$ (kHz) | -848.764 | -832.792 | -1566.435 | -889.960 | -235.645 | -294.363 | -187.054 | -22.582 | -1198.246 | -748.573 | -851.054 | -1194.729 |
| $|\mu_a|$ (D) | 0.51 | 0.52 | 0.57 | 0.47 | 0.52 | 0.50 | 0.43 | 0.38 | 0.47 | 0.47 | 0.47 | 0.48 |
| $|\mu_b|$ (D) | 1.44 | 1.49 | 1.42 | 1.42 | 1.67 | 1.49 | 1.51 | 1.51 | 1.35 | 1.51 | 1.42 | 1.53 |
| $|\mu_c|$ (D) | 0.00 | 0.00 | 0.00 | 0.00 | 0.00 | 0.00 | 0.00 | 0.00 | 0.00 | 0.00 | 0.00 | 0.00 |
| $|\mu_{TOT}|$ (D) | 1.53 | 1.58 | 1.53 | 1.50 | 1.75 | 1.58 | 1.57 | 1.56 | 1.43 | 1.58 | 1.49 | 1.60 |

**Table S2.8** Rotational constants, centrifugal distortion constants and dipole moment of BA-w calculated with twelve different methods

| BA-w | B2PLYP-D3(BJ) | B3LYP-D3(BJ) | BLYP-D3(BJ) | DSD-PBEP86-D3(BJ) | HF | M06-2X(D3) | MN15-D3(BJ) | MN15L(D3) | MP2 | PBE0-D3(BJ) | revDSD-PBEP86-D3(BJ) | ωB97X-D |
|---|---|---|---|---|---|---|---|---|---|---|---|---|
| $A$ (MHz) | 7829.5115 | 7844.2604 | 7689.5433 | 7826.0896 | 8016.4654 | 7864.9509 | 7846.3049 | 7681.3671 | 7815.3657 | 7913.3629 | 7815.0047 | 7897.2300 |
| $B$ (MHz) | 967.9590 | 965.9692 | 952.5757 | 970.7453 | 929.4303 | 970.3324 | 971.8073 | 934.5152 | 973.3060 | 980.8581 | 966.0161 | 965.3186 |
| $C$ (MHz) | 876.6702 | 875.2956 | 862.7138 | 879.0192 | 846.4811 | 878.8888 | 879.9548 | 847.5953 | 880.8855 | 888.4935 | 875.0137 | 875.4774 |
| $\Delta_J$ (kHz) | 78.189 | 76.439 | 76.250 | 78.881 | 101.802 | 78.687 | 82.338 | 90.470 | 81.704 | 77.463 | 80.363 | 78.836 |
| $\Delta_K$ (kHz) | 12346.944 | 12234.202 | 11592.211 | 12190.859 | 23293.460 | 13099.204 | 12657.127 | 15938.258 | 12264.943 | 11406.205 | 12648.301 | 12362.345 |
| $\Delta_{JK}$ (kHz) | -318.529 | -307.676 | -262.466 | -328.487 | -622.115 | -291.257 | -335.767 | -314.530 | -332.651 | -315.684 | -343.932 | -380.783 |
| $\delta_J$ (kHz) | 9.482 | 9.245 | 9.152 | 9.587 | 12.106 | 9.469 | 10.014 | 10.538 | 9.901 | 9.407 | 9.750 | 9.622 |
| $\delta_K$ (kHz) | -175.727 | -95.006 | -236.122 | -139.249 | 165.482 | 56.831 | 98.839 | 167.472 | -336.765 | -81.009 | -116.326 | -79.884 |
| $|\mu_a|$ (D) | 0.51 | 0.59 | 0.69 | 0.48 | 0.28 | 0.39 | 0.42 | 0.25 | 0.42 | 0.60 | 0.47 | 0.57 |
| $|\mu_b|$ (D) | 0.09 | 0.17 | 0.18 | 0.08 | 0.06 | 0.05 | 0.10 | 0.11 | 0.00 | 0.22 | 0.07 | 0.21 |

| | | | | | | | | | | | | |
|---|---|---|---|---|---|---|---|---|---|---|---|---|
| $|\mu_c|$ (D) | 1.22 | 1.24 | 1.27 | 1.23 | 1.02 | 1.16 | 1.20 | 1.15 | 1.21 | 1.27 | 1.22 | 1.26 |
| $|\mu_{TOT}|$ (D) | 1.32 | 1.38 | 1.46 | 1.32 | 1.06 | 1.22 | 1.27 | 1.18 | 1.28 | 1.42 | 1.31 | 1.40 |

**Table S2.9** Rotational constants, centrifugal distortion constants and dipole moment of BA-w$_2$ calculated with twelve different methods

| Parameters | B2PLYP-D3(BJ) | B3LYP-D3(BJ) | BLYP-D3(BJ) | DSD-PBEP86-D3(BJ) | HF | M06-2X(D3) | MN15-D3(BJ) | MN15L(D3) | MP2 | PBE0-D3(BJ) | revDSD-PBEP86-D3(BJ) | ωB97X-D |
|---|---|---|---|---|---|---|---|---|---|---|---|---|
| A (MHz) | 3592.9077 | 3624.1269 | 3598.6807 | 3579.1810 | 3370.1953 | 3541.4690 | 3502.6142 | 3228.7497 | 3565.7799 | 3667.8939 | 3559.6124 | 3614.4992 |
| B (MHz) | 694.4808 | 693.7836 | 683.0176 | 696.6179 | 672.9480 | 703.4655 | 698.6989 | 677.7988 | 698.0926 | 704.1223 | 693.5357 | 694.9802 |
| C (MHz) | 589.4336 | 589.8396 | 581.6484 | 590.6745 | 567.2414 | 594.4675 | 589.9072 | 567.3876 | 591.2764 | 598.5800 | 587.9198 | 590.5375 |
| $\Delta_J$ (kHz) | 39.464 | 37.688 | 36.325 | 40.584 | 55.022 | 45.740 | 47.280 | 62.239 | 41.992 | 38.298 | 41.187 | 39.087 |
| $\Delta_K$ (kHz) | 4799.566 | 4573.004 | 4240.510 | 4925.045 | 7136.393 | 5885.152 | 6001.041 | 8330.501 | 5051.887 | 4562.659 | 5001.265 | 4809.002 |
| $\Delta_{JK}$ (kHz) | 57.223 | 63.914 | 99.534 | 50.772 | -24.837 | -12.715 | -47.058 | -206.092 | 64.523 | 74.256 | 51.082 | 56.480 |
| $\delta_J$ (kHz) | 7.556 | 7.078 | 6.516 | 7.832 | 11.578 | 9.234 | 9.982 | 14.750 | 8.092 | 7.103 | 7.964 | 7.399 |
| $\delta_K$ (kHz) | 147.583 | 151.037 | 115.962 | 160.296 | 262.207 | 213.952 | 221.831 | 298.770 | 138.321 | 157.758 | 165.787 | 174.549 |
| $|\mu_a|$ (D) | 0.34 | 0.40 | 0.46 | 0.32 | 0.36 | 0.30 | 0.24 | 0.35 | 0.26 | 0.37 | 0.33 | 0.46 |
| $|\mu_b|$ (D) | 0.15 | 0.09 | 0.08 | 0.17 | 0.12 | 0.23 | 0.17 | 0.15 | 0.25 | 0.06 | 0.17 | 0.06 |
| $|\mu_c|$ (D) | 2.18 | 2.23 | 2.32 | 2.20 | 1.67 | 2.14 | 2.09 | 2.12 | 2.16 | 2.29 | 2.19 | 2.27 |
| $|\mu_{TOT}|$ (D) | 2.21 | 2.27 | 2.36 | 2.23 | 1.71 | 2.17 | 2.12 | 2.16 | 2.19 | 2.32 | 2.23 | 2.31 |

**Table S2.10** Rotational constants, centrifugal distortion constants and dipole moment of PTA calculated with twelve different methods

| Parameters | B2PLYP-D3(BJ) | B3LYP-D3(BJ) | BLYP-D3(BJ) | DSD-PBEP86-D3(BJ) | HF | M06-2X(D3) | MN15-D3(BJ) | MN15L(D3) | MP2 | PBE0-D3(BJ) | revDSD-PBEP86-D3(BJ) | ωB97X-D |
|---|---|---|---|---|---|---|---|---|---|---|---|---|
| A (MHz) | 7991.8900 | 8007.6913 | 7850.3819 | 7983.3497 | 8218.6240 | 8053.6516 | 8042.3390 | 7898.6245 | 7962.7461 | 8055.8492 | 7974.9854 | 8057.3717 |
| B (MHz) | 1052.1225 | 1048.1039 | 1030.7338 | 1055.3432 | 1052.4925 | 1058.6487 | 1062.0915 | 1048.3243 | 1059.0208 | 1058.8714 | 1053.1017 | 1053.1948 |
| C (MHz) | 951.2569 | 948.2069 | 932.0382 | 953.8800 | 954.4836 | 957.5086 | 960.1512 | 947.2912 | 956.5074 | 957.7153 | 951.9581 | 953.0833 |
| $\Delta_J$ (kHz) | 47.391 | 47.060 | 47.500 | 47.668 | 42.613 | 47.308 | 50.089 | 46.181 | 48.218 | 48.305 | 47.532 | 46.588 |
| $\Delta_K$ (kHz) | 3355.895 | 3361.243 | 3400.967 | 3347.380 | 3134.674 | 3368.872 | 3607.009 | 3338.498 | 3342.794 | 3381.046 | 3342.571 | 3333.752 |
| $\Delta_{JK}$ (kHz) | 505.321 | 495.893 | 518.034 | 498.136 | 423.434 | 461.301 | 484.374 | 450.997 | 509.937 | 507.290 | 497.260 | 475.481 |
| $\delta_J$ (kHz) | 5.102 | 5.031 | 5.071 | 5.163 | 4.461 | 5.120 | 5.498 | 5.063 | 5.261 | 5.216 | 5.141 | 4.988 |
| $\delta_K$ (kHz) | -433.436 | -290.248 | -622.151 | -330.921 | -105.416 | -30.951 | -48.363 | -35.492 | -450.542 | -278.512 | -339.199 | -169.411 |
| $|\mu_a|$ (D) | 0.39 | 0.41 | 0.47 | 0.35 | 0.39 | 0.39 | 0.31 | 0.26 | 0.36 | 0.35 | 0.35 | 0.36 |
| $|\mu_b|$ (D) | 1.54 | 1.59 | 1.53 | 1.52 | 1.77 | 1.59 | 1.60 | 1.62 | 1.44 | 1.61 | 1.51 | 1.62 |
| $|\mu_c|$ (D) | 0.00 | 0.00 | 0.00 | 0.00 | 0.00 | 0.00 | 0.00 | 0.00 | 0.00 | 0.00 | 0.00 | 0.00 |
| $|\mu_{TOT}|$ (D) | 1.59 | 1.64 | 1.60 | 1.56 | 1.81 | 1.63 | 1.63 | 1.64 | 1.49 | 1.65 | 1.55 | 1.66 |

**Table S2.11** Rotational constants, centrifugal distortion constants and dipole moment of PTA-w$_2$ calculated with twelve different methods

| Parameters | B2PLYP-D3(BJ) | B3LYP-D3(BJ) | BLYP-D3(BJ) | DSD-PBEP86-D3(BJ) | HF | M06-2X(D3) | MN15-D3(BJ) | MN15L(D3) | MP2 | PBE0-D3(BJ) | revDSD-PBEP86-D3(BJ) | ωB97X-D |
|---|---|---|---|---|---|---|---|---|---|---|---|---|
| A (MHz) | 3091.2893 | 3123.6539 | 3111.1456 | 3073.1496 | 2892.1505 | 3020.5256 | 2987.2235 | 2736.2012 | 3057.7976 | 3158.3475 | 3056.5061 | 3111.8644 |
| B (MHz) | 479.0538 | 477.9187 | 469.9829 | 480.9284 | 468.0973 | 486.1966 | 484.1985 | 473.1926 | 482.3601 | 484.6365 | 479.0341 | 479.2826 |
| C (MHz) | 419.6253 | 419.3825 | 413.1947 | 420.7689 | 407.1309 | 423.7430 | 421.5228 | 408.1616 | 421.5325 | 425.2377 | 419.0049 | 420.2871 |
| $\Delta_J$ (kHz) | 20.586 | 19.583 | 18.752 | 21.256 | 27.157 | 24.076 | 25.097 | 32.439 | 22.082 | 20.058 | 21.448 | 20.369 |

| Parameters | B2PLYP-D3(BJ) | B3LYP-D3(BJ) | BLYP-D3(BJ) | DSD-PBEP86-D3(BJ) | HF | M06-2X(D3) | MN15-D3(BJ) | MN15L(D3) | MP2 | PBE0-D3(BJ) | revDSD-PBEP86-D3(BJ) | ωB97X-D |
|---|---|---|---|---|---|---|---|---|---|---|---|---|
| $\Delta_K$ (kHz) | 5945.184 | 5729.936 | 5443.093 | 6048.377 | 7839.400 | 7094.152 | 6965.063 | 8732.798 | 6201.762 | 5761.545 | 6096.127 | 5977.159 |
| $\Delta_{JK}$ (kHz) | -224.362 | -209.585 | -178.256 | -235.251 | -317.721 | -308.197 | -328.230 | -473.490 | -236.319 | -210.583 | -235.032 | -219.037 |
| $\delta_J$ (kHz) | 4.248 | 3.964 | 3.670 | 4.430 | 6.010 | 5.344 | 5.635 | 8.134 | 4.629 | 4.038 | 4.470 | 4.165 |
| $\delta_K$ (kHz) | 69.185 | 74.013 | 59.697 | 76.460 | 123.333 | 101.569 | 108.698 | 143.998 | 54.572 | 76.901 | 79.791 | 54.007 |
| $|\mu_a|$ (D) | 0.41 | 0.47 | 0.53 | 0.39 | 0.42 | 0.37 | 0.31 | 0.41 | 0.34 | 0.43 | 0.40 | 0.52 |
| $|\mu_b|$ (D) | 0.07 | 0.00 | 0.02 | 0.09 | 0.02 | 0.15 | 0.08 | 0.03 | 0.17 | 0.03 | 0.09 | 0.02 |
| $|\mu_c|$ (D) | 2.17 | 2.23 | 2.30 | 2.19 | 1.66 | 2.12 | 2.09 | 2.10 | 2.15 | 2.29 | 2.19 | 2.25 |
| $|\mu_{TOT}|$ (D) | 2.21 | 2.27 | 2.37 | 2.23 | 1.72 | 2.15 | 2.11 | 2.14 | 2.18 | 2.33 | 2.22 | 2.31 |

**Table S2.12** Rotational constants, centrifugal distortion constants and dipole moment of HXA calculated with twelve different methods

| Parameters | B2PLYP-D3(BJ) | B3LYP-D3(BJ) | BLYP-D3(BJ) | DSD-PBEP86-D3(BJ) | HF | M06-2X(D3) | MN15-D3(BJ) | MN15L(D3) | MP2 | PBE0-D3(BJ) | revDSD-PBEP86-D3(BJ) | ωB97X-D |
|---|---|---|---|---|---|---|---|---|---|---|---|---|
| $A$ (MHz) | 7420.5676 | 7432.6209 | 7287.1353 | 7413.8716 | 7617.6190 | 7480.1767 | 7471.6472 | 7335.6920 | 7398.4917 | 7478.1352 | 7404.9191 | 7476.3561 |
| $B$ (MHz) | 651.5545 | 649.1713 | 638.6964 | 653.4767 | 651.5206 | 655.0984 | 657.3631 | 648.3649 | 655.4775 | 655.7523 | 651.9489 | 652.4109 |
| $C$ (MHz) | 610.0886 | 608.0852 | 598.0515 | 611.7872 | 611.2455 | 613.6205 | 615.5446 | 606.9627 | 613.3958 | 614.1785 | 610.4004 | 611.2282 |
| $\Delta_J$ (kHz) | 14.238 | 14.209 | 14.405 | 14.260 | 12.885 | 14.199 | 14.856 | 13.687 | 14.368 | 14.445 | 14.228 | 14.044 |
| $\Delta_K$ (kHz) | 2602.914 | 2610.049 | 2696.710 | 2571.260 | 2343.550 | 2540.300 | 2691.906 | 2485.311 | 2596.983 | 2577.176 | 2568.083 | 2553.751 |
| $\Delta_{JK}$ (kHz) | 306.312 | 299.093 | 304.499 | 307.581 | 258.956 | 293.836 | 317.763 | 295.723 | 312.688 | 313.657 | 304.687 | 292.265 |
| $\delta_J$ (kHz) | 0.912 | 0.907 | 0.924 | 0.915 | 0.803 | 0.908 | 0.957 | 0.880 | 0.928 | 0.923 | 0.913 | 0.896 |
| $\delta_K$ (kHz) | -272.959 | -214.600 | -370.263 | -234.898 | -110.476 | -68.797 | -92.539 | -58.628 | -235.016 | -201.212 | -213.837 | -201.842 |
| $|\mu_a|$ (D) | 0.49 | 0.51 | 0.57 | 0.45 | 0.49 | 0.48 | 0.40 | 0.36 | 0.46 | 0.45 | 0.45 | 0.46 |
| $|\mu_b|$ (D) | 1.43 | 1.48 | 1.41 | 1.41 | 1.66 | 1.48 | 1.49 | 1.48 | 1.34 | 1.50 | 1.40 | 1.51 |
| $|\mu_c|$ (D) | 0.00 | 0.00 | 0.00 | 0.00 | 0.00 | 0.00 | 0.00 | 0.00 | 0.00 | 0.00 | 0.00 | 0.00 |
| $|\mu_{TOT}|$ (D) | 1.51 | 1.56 | 1.52 | 1.48 | 1.73 | 1.56 | 1.54 | 1.53 | 1.41 | 1.56 | 1.48 | 1.58 |

**Table S2.13** Rotational constants, centrifugal distortion constants and dipole moment of HXA-w calculated with twelve different methods

| Parameters | B2PLYP-D3(BJ) | B3LYP-D3(BJ) | BLYP-D3(BJ) | DSD-PBEP86-D3(BJ) | HF | M06-2X(D3) | MN15-D3(BJ) | MN15L(D3) | MP2 | PBE0-D3(BJ) | revDSD-PBEP86-D3(BJ) | ωB97X-D |
|---|---|---|---|---|---|---|---|---|---|---|---|---|
| $A$ (MHz) | 6063.3387 | 6076.3604 | 5975.2623 | 6053.1419 | 6139.2430 | 6058.9487 | 6040.2879 | 5891.3972 | 6051.7422 | 6127.8848 | 6043.3565 | 6109.1446 |
| $B$ (MHz) | 426.0936 | 424.9325 | 418.6466 | 427.3358 | 415.8155 | 427.9117 | 429.0396 | 416.7897 | 428.6385 | 430.5302 | 425.7664 | 425.5990 |
| $C$ (MHz) | 403.3137 | 402.3455 | 396.3619 | 404.4163 | 394.2585 | 404.8887 | 405.8328 | 394.3126 | 405.5511 | 407.6157 | 402.9707 | 403.0990 |
| $\Delta_J$ (kHz) | 10.040 | 9.870 | 9.856 | 10.139 | 11.645 | 10.170 | 10.687 | 10.942 | 10.374 | 10.075 | 10.235 | 9.973 |
| $\Delta_K$ (kHz) | 17345.106 | 17168.547 | 16435.096 | 17269.169 | 27817.756 | 18508.331 | 18541.633 | 21567.944 | 17365.829 | 16682.440 | 17685.174 | 17100.083 |
| $\Delta_{JK}$ (kHz) | -276.057 | -270.960 | -252.506 | -279.968 | -394.304 | -283.523 | -301.982 | -302.382 | -282.070 | -277.078 | -285.471 | -284.343 |
| $\delta_J$ (kHz) | 0.965 | 0.945 | 0.928 | 0.978 | 1.132 | 0.994 | 1.057 | 1.053 | 0.997 | 0.972 | 0.988 | 0.961 |
| $\delta_K$ (kHz) | -72.002 | -51.818 | -112.793 | -53.805 | 14.000 | -64.294 | 0.529 | 19.888 | -110.569 | -48.101 | -50.547 | -108.253 |
| $|\mu_a|$ (D) | 0.56 | 0.64 | 0.74 | 0.53 | 0.33 | 0.44 | 0.46 | 0.28 | 0.47 | 0.64 | 0.51 | 0.61 |
| $|\mu_b|$ (D) | 0.11 | 0.19 | 0.22 | 0.10 | 0.06 | 0.05 | 0.11 | 0.10 | 0.01 | 0.24 | 0.09 | 0.23 |
| $|\mu_c|$ (D) | 1.21 | 1.24 | 1.26 | 1.23 | 1.02 | 1.16 | 1.19 | 1.15 | 1.20 | 1.27 | 1.22 | 1.26 |
| $|\mu_{TOT}|$ (D) | 1.34 | 1.40 | 1.48 | 1.34 | 1.08 | 1.24 | 1.28 | 1.18 | 1.29 | 1.44 | 1.33 | 1.41 |

**Table S2.14** Rotational constants, centrifugal distortion constants and dipole moment of HXA-w$_2$ calculated with twelve different methods

| Parameters | B2PLYP-D3(BJ) | B3LYP-D3(BJ) | BLYP-D3(BJ) | DSD-PBEP86-D3(BJ) | HF | M06-2X(D3) | MN15-D3(BJ) | MN15L(D3) | MP2 | PBE0-D3(BJ) | revDSD-PBEP86-D3(BJ) | ωB97X-D |
|---|---|---|---|---|---|---|---|---|---|---|---|---|
| *A* (MHz) | 3029.1023 | 3062.5116 | 3050.5958 | 3011.4970 | 2831.8023 | 2956.9422 | 2925.3332 | 2675.2749 | 2997.3882 | 3094.5562 | 2995.0238 | 3048.3936 |
| *B* (MHz) | 330.3368 | 329.4451 | 324.0266 | 331.5228 | 324.2886 | 334.6521 | 333.7884 | 326.8797 | 332.5010 | 333.8362 | 330.3410 | 330.5864 |
| *C* (MHz) | 300.8998 | 300.5021 | 295.9637 | 301.7329 | 293.6923 | 303.7347 | 302.6673 | 294.2963 | 302.3796 | 304.4946 | 300.5876 | 301.3531 |
| $\Delta_J$ (kHz) | 7.716 | 7.358 | 7.057 | 7.948 | 10.040 | 8.981 | 9.402 | 11.897 | 8.236 | 7.526 | 8.020 | 7.545 |
| $\Delta_K$ (kHz) | 6197.385 | 5966.631 | 5645.396 | 6314.753 | 8128.109 | 7459.825 | 7350.997 | 9069.936 | 6492.964 | 6012.564 | 6365.046 | 6146.245 |
| $\Delta_{JK}$ (kHz) | -132.021 | -122.298 | -101.945 | -138.595 | -196.582 | -188.376 | -198.099 | -285.153 | -139.576 | -122.409 | -139.026 | -124.300 |
| $\delta_J$ (kHz) | 1.277 | 1.189 | 1.095 | 1.331 | 1.823 | 1.635 | 1.722 | 2.486 | 1.388 | 1.211 | 1.345 | 1.234 |
| $\delta_K$ (kHz) | 36.807 | 39.500 | 27.278 | 41.621 | 69.488 | 54.899 | 65.403 | 83.988 | 20.391 | 45.374 | 43.284 | 7.836 |
| $|\mu_a|$ (D) | 0.41 | 0.47 | 0.53 | 0.40 | 0.43 | 0.37 | 0.31 | 0.41 | 0.34 | 0.44 | 0.40 | 0.52 |
| $|\mu_b|$ (D) | 0.15 | 0.08 | 0.07 | 0.17 | 0.11 | 0.23 | 0.17 | 0.13 | 0.25 | 0.05 | 0.17 | 0.05 |
| $|\mu_c|$ (D) | 2.18 | 2.23 | 2.31 | 2.20 | 1.67 | 2.12 | 2.09 | 2.12 | 2.15 | 2.29 | 2.19 | 2.26 |
| $|\mu_{TOT}|$ (D) | 2.22 | 2.28 | 2.37 | 2.24 | 1.73 | 2.16 | 2.12 | 2.16 | 2.19 | 2.33 | 2.23 | 2.32 |

## 4. Benchmark data of evaluating the optimal method for calculating rotational constants

**Figure S1** The absolute errors for rotational constant *B* of AA, AA-w, AA-w$_2$, PPA, PPA-w, and PPA-w$_2$ by using twelve methods (may-cc-pVTZ basis set)

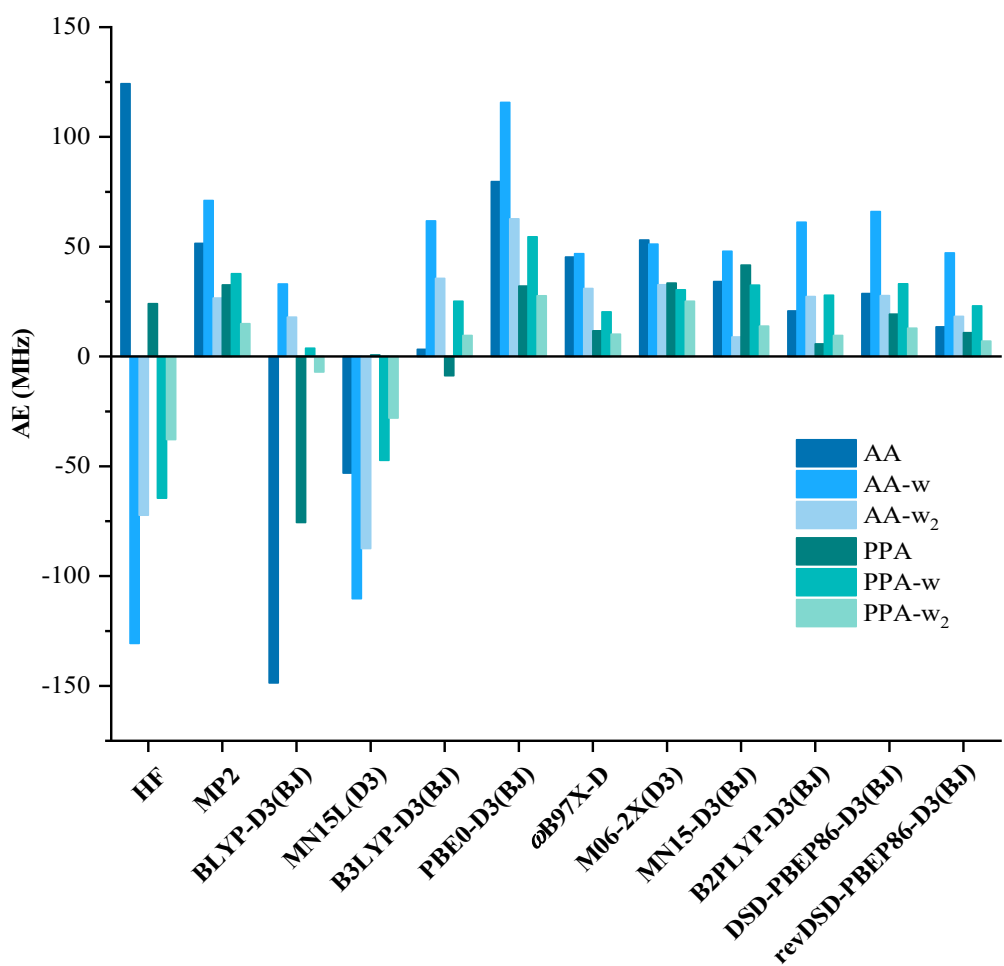

**Figure S2** The absolute errors for rotational constant *C* of AA, AA-w, AA-w$_2$, PPA, PPA-w, and PPA-w$_2$ by using twelve methods (may-cc-pVTZ basis set)

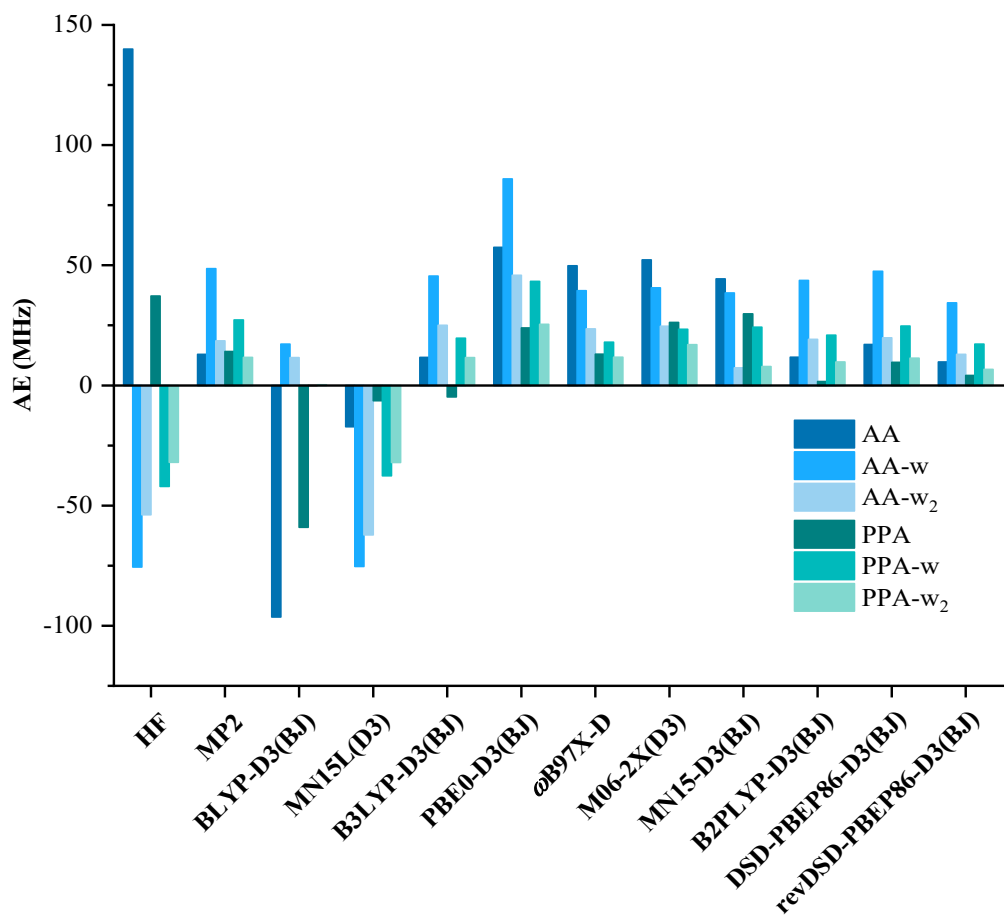

**Table S3** Systematic error of calculating rotational constants with twelve different methods

|  | B2PLYP-D3(BJ) | B3LYP-D3(BJ) | BLYP-D3(BJ) | DSD-PBEP86-D3(BJ) | HF | M06-2X(D3) | MN15-D3(BJ) | MN15L(D3) | MP2 | PBE0-D3(BJ) | revDSD-PBEP86-D3(BJ) | ωB97X-D |
|---|---|---|---|---|---|---|---|---|---|---|---|---|
| A(slope) | 0.98984 | 0.98992 | 0.95905 | 0.99123 | 1.08079 | 1.00871 | 1.01366 | 1.02546 | 0.98709 | 0.99396 | 0.9926 | 1.00406 |
| B(slope) | 0.99925 | 0.99675 | 0.97996 | 0.99992 | 1.02424 | 1.00266 | 1.00173 | 1.00125 | 1.00257 | 1.00288 | 0.99922 | 1.00267 |
| C(slope) | 0.99818 | 0.99729 | 0.97492 | 0.99917 | 1.04408 | 1.0069 | 1.00759 | 1.00771 | 0.99833 | 1.00349 | 0.99903 | 1.0068 |
| MAD | 0.00424 | 0.00535 | 0.02869 | 0.00323 | 0.04970 | 0.00609 | 0.00766 | 0.01147 | 0.00572 | 0.00414 | 0.00305 | 0.00451 |
| A(intercept) | 108.40073 | 129.45978 | 207.90388 | 96.60279 | -464.3187 | 22.18935 | -43.40296 | -340.1455 | 105.00327 | 171.91164 | 70.97013 | 71.37058 |
| B(intercept) | 27.94642 | 32.06445 | 38.12499 | 31.56615 | -107.8595 | 28.69889 | 24.0072 | -58.44224 | 30.44168 | 52.36618 | 22.60308 | 18.58634 |
| C(intercept) | 21.99836 | 24.34172 | 36.43542 | 23.5586 | -105.4792 | 14.85106 | 7.9139 | -56.16142 | 26.0097 | 38.98884 | 16.38501 | 10.33543 |
| MAD | 52.781837 | 61.955317 | 94.154763 | 50.575847 | 225.8858 | 21.9131 | 25.10802 | 151.58305 | 53.818217 | 87.755553 | 36.65274 | 33.430783 |
| A(R$^2$) | 0.99998 | 0.99998 | 0.99995 | 0.99999 | 0.99989 | 0.99995 | 0.99996 | 0.99997 | 0.99999 | 0.99999 | 0.99999 | 0.99998 |
| B(R$^2$) | 0.99996 | 0.99994 | 0.99991 | 0.99997 | 0.99979 | 0.99999 | 0.99998 | 0.99984 | 0.99997 | 0.9999 | 0.99998 | 0.99998 |
| C(R$^2$) | 0.99993 | 0.9999 | 0.99985 | 0.99993 | 0.99962 | 0.99998 | 0.99997 | 0.99981 | 0.99993 | 0.99982 | 0.99996 | 0.99996 |

All of the results can show strong linear correlation. We consider the slope that is closer to 1 and has smaller error as the main dominance, and the intercept close to 0 is the auxiliary reference.

## 5. SAPT analysis

**Table S4** SAPT energy of electrostatics, exchange, induction, dispersion and total SAPT2+(3) δMP2 energy

| Electronic energy (kcal/mol) | Electrostatics | Exchange | Induction | Dispersion | Total SAPT2+(3)dMP2 |
|---|---|---|---|---|---|
| AA-w | -21.3 | 24.8 | -8.0 | -6.5 | -10.7 |
| AA-w$_2$ (i) | -29.4 | 32.3 | -13.4 | -8.3 | -18.8 |
| AA-w$_2$ (ii) | -23.7 | 27.0 | -10.2 | -7.1 | -14.0 |
| AA-w$_2$ (iii) | -28.5 | 33.6 | -13.6 | -8.1 | -16.6 |
| PPA-w | -20.9 | 24.8 | -7.9 | -6.6 | -10.6 |
| PPA-w$_2$ (i) | -29.2 | 32.1 | -13.3 | -8.3 | -18.7 |
| PPA-w$_2$ (ii) | -23.7 | 26.9 | -10.1 | -7.1 | -14.0 |
| PPA-w$_2$ (iii) | -28.3 | 33.3 | -13.4 | -8.1 | -16.5 |
| BA-w | -20.9 | 24.8 | -7.9 | -6.6 | -10.6 |
| BA-w$_2$ (i) | -29.1 | 32.1 | -13.2 | -8.4 | -18.7 |
| BA-w$_2$ (ii) | -23.7 | 26.9 | -10.1 | -7.1 | -14.0 |
| BA-w$_2$ (iii) | -28.2 | 33.2 | -13.3 | -8.1 | -16.4 |
| PTA-w | -20.9 | 24.8 | -7.9 | -6.6 | -10.6 |
| PTA-w$_2$ (i) | -29.1 | 32.1 | -13.2 | -8.4 | -18.7 |
| PTA-w$_2$ (ii) | -23.7 | 27.0 | -10.2 | -7.1 | -14.0 |
| PTA-w$_2$ (iii) | -28.2 | 33.2 | -13.3 | -8.1 | -16.4 |
| HXA-w | -20.9 | 24.8 | -7.9 | -6.6 | -10.6 |
| HXA-w$_2$ (i) | -29.1 | 32.1 | -13.2 | -8.4 | -18.7 |
| HXA-w$_2$ (ii) | -23.7 | 27.0 | -10.2 | -7.1 | -14.0 |
| HXA-w$_2$ (iii) | -28.2 | 33.2 | -13.3 | -8.1 | -16.4 |

**Table S5** Energy (Thermo correction + electronic energy) of the substances under different temperature

| G(A.U.) | 0 | 2 K | 4 K | 5 K | 6 K | 8 K | 10 K | 30 K | 50 K | 100 K | 150 K | 200 K | 273K | 298 K | 350 K |
|---|---|---|---|---|---|---|---|---|---|---|---|---|---|---|---|
| AA | -228.733 | -228.733 | -228.733 | -228.733 | -228.733 | -228.733 | -228.733 | -228.734 | -228.736 | -228.74 | -228.745 | -228.749 | -228.757 | -228.76 | -228.766 |
| AA-w | -305.068 | -305.068 | -305.068 | -305.068 | -305.068 | -305.068 | -305.069 | -305.07 | -305.072 | -305.076 | -305.081 | -305.087 | -305.096 | -305.099 | -305.106 |
| AA-w$_2$ | -381.405 | -381.405 | -381.405 | -381.405 | -381.405 | -381.405 | -381.405 | -381.407 | -381.408 | -381.413 | -381.419 | -381.426 | -381.436 | -381.44 | -381.449 |
| PPA | -267.952 | -267.952 | -267.952 | -267.952 | -267.953 | -267.953 | -267.953 | -267.954 | -267.956 | -267.96 | -267.965 | -267.97 | -267.979 | -267.982 | -267.988 |
| PPA-w | -344.288 | -344.288 | -344.288 | -344.288 | -344.288 | -344.288 | -344.288 | -344.29 | -344.291 | -344.296 | -344.302 | -344.308 | -344.317 | -344.321 | -344.329 |
| PPA-w$_2$ | -420.624 | -420.624 | -420.624 | -420.624 | -420.625 | -420.625 | -420.625 | -420.626 | -420.628 | -420.633 | -420.64 | -420.647 | -420.658 | -420.662 | -420.671 |
| BA | -307.172 | -307.172 | -307.172 | -307.172 | -307.172 | -307.172 | -307.172 | -307.174 | -307.175 | -307.18 | -307.185 | -307.191 | -307.2 | -307.203 | -307.211 |
| BA-w | -383.507 | -383.507 | -383.507 | -383.507 | -383.508 | -383.508 | -383.508 | -383.509 | -383.511 | -383.516 | -383.522 | -383.529 | -383.539 | -383.543 | -383.551 |
| BA-w$_2$ | -459.844 | -459.844 | -459.844 | -459.844 | -459.844 | -459.844 | -459.844 | -459.846 | -459.848 | -459.853 | -459.86 | -459.867 | -459.879 | -459.884 | -459.893 |
| PTA | -346.392 | -346.392 | -346.392 | -346.392 | -346.392 | -346.392 | -346.392 | -346.394 | -346.395 | -346.4 | -346.406 | -346.412 | -346.422 | -346.426 | -346.433 |
| PTA-w | -422.727 | -422.727 | -422.727 | -422.727 | -422.727 | -422.727 | -422.728 | -422.729 | -422.731 | -422.736 | -422.743 | -422.75 | -422.761 | -422.765 | -422.774 |
| PTA-w$_2$ | -499.064 | -499.064 | -499.064 | -499.064 | -499.064 | -499.064 | -499.064 | -499.066 | -499.068 | -499.074 | -499.081 | -499.088 | -499.101 | -499.106 | -499.116 |
| HXA | -385.611 | -385.611 | -385.612 | -385.612 | -385.612 | -385.612 | -385.612 | -385.614 | -385.615 | -385.621 | -385.627 | -385.633 | -385.644 | -385.648 | -385.656 |
| HXA-w | -461.947 | -461.947 | -461.947 | -461.947 | -461.947 | -461.947 | -461.947 | -461.949 | -461.951 | -461.957 | -461.963 | -461.971 | -461.983 | -461.987 | -461.997 |
| HXA-w$_2$ | -538.283 | -538.283 | -538.284 | -538.284 | -538.284 | -538.284 | -538.284 | -538.286 | -538.288 | -538.294 | -538.301 | -538.309 | -538.323 | -538.328 | -538.339 |
| H2O | -76.3229 | -76.3228 | -76.3229 | -76.3229 | -76.3229 | -76.323 | -76.323 | -76.3238 | -76.3247 | -76.3274 | -76.3304 | -76.3337 | -76.3387 | -76.3405 | -76.3443 |

**Table S6** ΔG of the in the water-losing process at different tempertures

| ΔG | 0 | 2 K | 4 K | 5 K | 6 K | 8 K | 10 K | 30 K | 50 K | 100 K | 150 K | 200 K | 273 K | 298 K | 350 K |
|---|---|---|---|---|---|---|---|---|---|---|---|---|---|---|---|
| AA-w | 7.928 | 7.939 | 7.928 | 7.920 | 7.909 | 7.885 | 7.856 | 7.454 | 6.947 | 5.506 | 3.978 | 2.427 | 0.158 | -0.615 | -2.208 |
| AA-w$_2$ | 8.573 | 8.583 | 8.572 | 8.563 | 8.552 | 8.527 | 8.497 | 8.096 | 7.603 | 6.223 | 4.761 | 3.274 | 1.092 | 0.349 | -1.185 |
| PPA-w | 7.890 | 7.900 | 7.888 | 7.879 | 7.868 | 7.843 | 7.813 | 7.400 | 6.885 | 5.434 | 3.899 | 2.343 | 0.066 | -0.709 | -2.308 |
| PPA-w$_2$ | 8.568 | 8.578 | 8.565 | 8.556 | 8.545 | 8.518 | 8.489 | 8.081 | 7.583 | 6.193 | 4.720 | 3.222 | 1.026 | 0.278 | -1.266 |
| BA-w | 7.889 | 7.899 | 7.886 | 7.876 | 7.865 | 7.838 | 7.807 | 7.387 | 6.868 | 5.413 | 3.876 | 2.318 | 0.037 | -0.739 | -2.341 |
| BA-w$_2$ | 8.575 | 8.584 | 8.571 | 8.561 | 8.550 | 8.523 | 8.493 | 8.084 | 7.586 | 6.197 | 4.726 | 3.231 | 1.037 | 0.290 | -1.251 |
| PTA-w | 7.909 | 7.917 | 7.904 | 7.894 | 7.883 | 7.855 | 7.825 | 7.404 | 6.887 | 5.439 | 3.911 | 2.362 | 0.095 | -0.676 | -2.268 |
| PTA-w$_2$ | 8.571 | 8.580 | 8.567 | 8.557 | 8.545 | 8.518 | 8.488 | 8.076 | 7.576 | 6.181 | 4.705 | 3.203 | 1.001 | 0.251 | -1.296 |
| HXA-w | 7.912 | 7.920 | 7.906 | 7.896 | 7.884 | 7.857 | 7.825 | 7.403 | 6.886 | 5.440 | 3.913 | 2.365 | 0.100 | -0.670 | -2.261 |
| HXA-w$_2$ | 8.573 | 8.581 | 8.568 | 8.558 | 8.546 | 8.519 | 8.488 | 8.075 | 7.575 | 6.179 | 4.701 | 3.198 | 0.994 | 0.244 | -1.305 |

## 6. AIMD simulations for dihedral angle of the terminal methyl

**Fig S3.1** Change of dihedral angle of the methyl of HXA-w in 10 ps (5K)

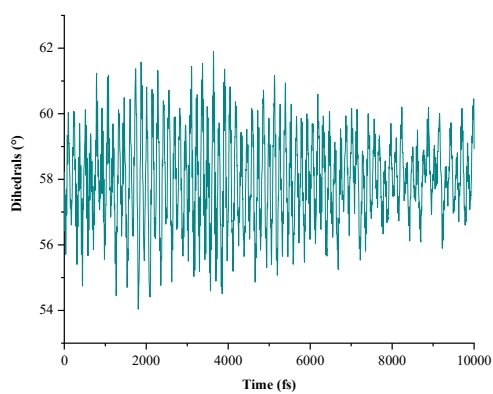

The dihedral angle of the methyl of the most stable structure is 58.01° and it varies from 54.03° to 61.90°.

**Fig S3.2** Change of dihedral angle of the methyl of BA-w in 10 ps (50K)

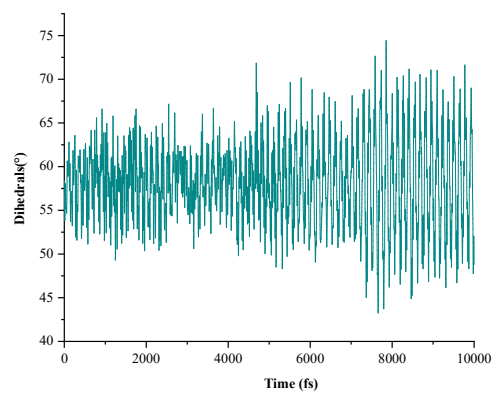

**Fig S3.3** Change of dihedral angle of the methyl of BA-w in 10 ps (100K)

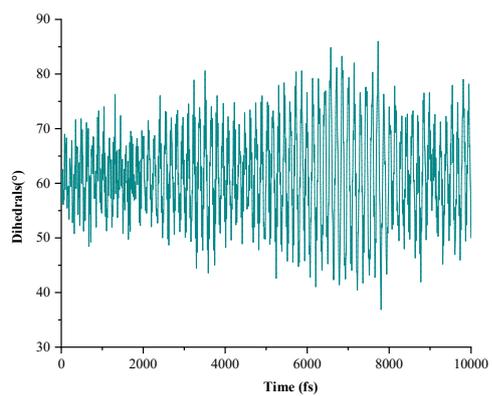

**Fig S3.4** Change of dihedral angle of the methyl of HXA-w in 10 ps (50K)

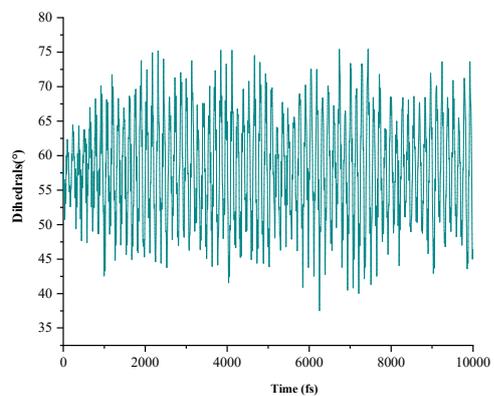

**Fig S3.5** Change of dihedral angle of the methyl of HXA-w in 10 ps (100K)

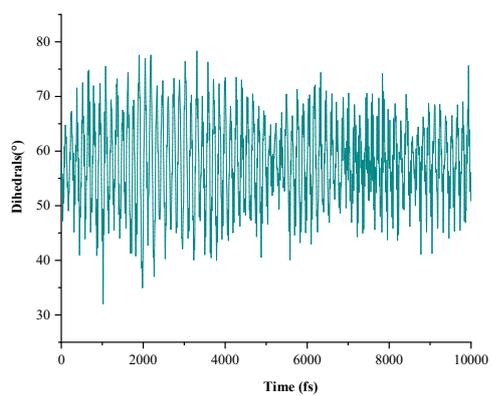